\documentclass{tglat2e}
\theoremstyle{plain}
\newtheorem{theorem}{Theorem}
\newtheorem{proposition}[theorem]{Proposition}
\newtheorem{corollary}[theorem]{Corollary}
\newtheorem{lemma}[theorem]{Lemma}

\theoremstyle{definition}
\newtheorem{definition}{Definition}
\newtheorem{example}{Example}

\theoremstyle{remark}
\newtheorem{remark}{Remark}

\begin{document}
\title[Combinatorics of Coinvariants]
{Combinatorics \\ of the $\widehat{\mathfrak{sl}}_2$ Spaces of Coinvariants}
\authors{B.~Feigin\thanks{Supported by grants RFBR 99-01-01169 and
INTAS-OPEN-97-1312}
\address  Landau institute\\for Theoretical Physics\\
Chernogolovka, 142432, Russia\email feigin@mccme.ru\and
R.~Kedem\thanks{Supported by NSF grant number 9870550}
\address  Dept. of Mathematics\\University of Massachusetts\\
Amherst, MA 01003, USA\email rinat@math.berkeley.edu
\and S.~Loktev\thanks{Partially supported by CRDF grant RM1-265}
\address Independent University of Moscow\\B. Vlasievsky per, 11\\
Moscow, Russia\email loktev@mccme.ru
\and T.~Miwa\address Department of Mathematics\\Kyoto University\\
Kyoto, Japan\email tetsuji@kusm.kyoto-u.ac.jp
\and E.~Mukhin\address Dept. of Mathematics\\
Univeristy of California\\ Berkeley, CA 94720, USA
\email mukhin@math.berkeley.edu
}
\dedicatory{Dedicated to the memory of Denis Uglov}
\maketitle
\begin{abstract}
We consider two types of quotients of the integrable modules of
$\widehat{\mathfrak{sl}}_2$.
These spaces of coinvariants have dimensions described
in terms of the Verlinde algebra of level $k$. We describe monomial bases for
the spaces of coinvariants, which leads to a fermionic description of
these spaces. For $k=1$, we give explicit formulas for the
characters. We also present recursion relations satisfied by the
characters and the monomial bases.
\end{abstract}
\newcommand{\bea}{\begin{eqnarray}}
\newcommand{\ena}{\end{eqnarray}}
\newcommand{\bean}{\begin{eqnarray*}}
\newcommand{\eean}{\end{eqnarray*}}
\newcommand{\Ref}[1]{{$($\ref{#1}$)$}}
\newcommand{\nc}{\newcommand}
\nc{\C}{{\Bbb C}}
\nc{\slt}{{\widehat{\mathfrak{sl}}_2}}
\newcommand{\be}{\begin{displaymath}}
\newcommand{\ee}{\end{displaymath}}
\newcommand{\res}{{\rm res}}
\nc{\notag}{\nonumber}
\newcommand{\Z}{{\Bbb Z}} 
\nc{\verm}{M_{k,l}}
\nc{\inte}{{L_{k,l}}}
\nc{\Vk}{{\cal V}_k}
\nc{\pkln}{{\cal P}_{k,l}^{(N)}}
\nc{\ckln}{{c_{k,l}^{(N)}}}
\def\qbin[#1;#2]{{\Bigl[{#1\atop#2}\Bigr]}}
\def \proj {\C P^1}
\def \slg {\mathfrak{sl}_2}
\def \tp {{\tilde{\cal P}}}
\def \CC {{\Bbb C}}
\def \bm{\hat{\frak n}_-}
\def \wsl {{\widehat{\mathfrak{sl}}_2}}
\def \wsld {{\mathfrak{sl}_2 \otimes \CC[t,t^{-1}]]}}
\def\b{{\frak b}}

\section{Introduction}
Let $\wsl$ be the central extension of the current algebra 
\bea
\wsl&=&\mathfrak{sl}_2\otimes \C[t,t^{-1}]]\oplus\C c\oplus\C d.
\ena
Here, $t$ is the coordinate at infinity, see Section \ref{definitions},
and by $\C[t,t^{-1}]]$ we mean the completion of the ring of Laurent
polynomials with respect to $t^{-1}$.

Let $\bm=\mathfrak{sl}_2\otimes
t^{-1}\C[[t^{-1}]]\oplus e\otimes1 \subset\wsl$ be the Lie
subalgebra generated by the annihilation operators, and let $\b\subset\wsl$ 
be a Lie subalgebra such that
\bea
\dim (\wsl/(\bm+\b))<\infty.
\ena
Let $L$ be an integrable representation of $\widehat{\mathfrak{sl}}_2$.
Define the space of coinvariants, $I(L,\b) = L/\b L$. It
is easy to show that this space is finite-dimensional.
In conformal field theory, spaces of coinvariants are
studied  in connection with modular functors, spaces of conformal blocks, etc.
(see \cite{F},\cite{S},\cite{TUY}).

The first natural question is how to find the dimension of $I(L,\b)$. For
general $\b$, the dimension is not known, but in some cases, 
Verlinde's rule 
gives us an answer. 

\def \ve {{\cal E}}
Verlinde's rule \cite{V} was formulated for the
subalgebra $\b$ associated with a compact algebraic curve and a point on it. 
Let $\ve$ be an algebraic curve, $P\in\ve$ a point and $\xi$ a rank-$2$
$SL_2$-bundle on $\ve$. Fix a local coordinate $t^{-1}$ and a trivialization
of the bundle in a small vicinity of $P$. These data determine a subalgebra
$\b(\ve,\xi) \subset \wsld$. Namely,
using the local coordinate and the trivialization of the bundle, we
can identify the algebra $\wsl$ with the algebra of infinitesimal
automorphisms (i.e., the $\mathfrak{sl}_2$
action induced from the $SL_2$ action)
of the bundle $\xi$ restricted to the punctured vicinity of $P$.
Let $\b(\ve,\xi)$ be the algebra of infinitesimal automorphhisms
of the bundle $\xi$ restricted to the domain $\ve \setminus P$.
Clearly, we have the natural inclusion $\b(\ve,\xi) \hookrightarrow \wsl$.
Classically, Verlinde's rule gives the dimension of the space of coinvariants
with respect to this subalgebra by reducing the calculation to
inductive steps governed by the Verlinde algebra.

\def \nsub {{\b(\ve,\xi; z_1, \dots, z_s; w_1, \dots, w_r)}}

Let us consider a somewhat more general situation.
Let $(z_1, \dots, z_s; w_1, \dots, w_r)$ be a set of distinct points,
not equal to $P$. At each point $w_i$, fix a line $l_i$ in the $2$-dimensional
fiber $\xi|_{w_i}$. Now define a subalgebra $\nsub\subset\b(\ve,\xi)$
of infinitesimal automorphism $g$ such that $g(z_i) = 0$ and
$g(w_i) |_{l_i} = 0$ for all $i$. In the next section we formulate
Verlinde's rule in the case when $\ve \cong \CC P^1$, $\xi$ is the trivial
bundle and either $s$ or $r$ is equal to zero.

We described a family of subalgebras in $\wsl$ depending on some geometric
data (curve, bundle, points, lines). It is known that the spaces $I(L,
\nsub)$ form a vector bundle over the moduli space of these data. Actually
this bundle is equipped with a projective connection and many other
interesting structures \cite{TUY}.

Above, we assumed that the moduli parameters $z_i$ and $w_i$ are distinct.
Consider a compactification of the moduli space. 
The simplest ``strata at infinity" arise if some points collide. With any point
of these strata, we can associate a Lie subalgebra in $\wsl$, so    
we can consider the space of coinvariants, i.e., the quotient space
with respect to this subalgebra. In general, it is possible to prove
only that the dimension of this space is still finite and
not less than the dimension at a generic point of the moduli space.

In our special case ($\ve = \CC P^1$, and $\xi$ is trivial), we conjecture that
the spaces of coinvariants form a vector bundle over the configuration space 
of points on $\CC P^1$. This means that the algebra $\nsub$ is well-defined
for all values of $z_i$, $w_i$, and the dimension of the space of
coinvariants does not depend on these values. In this paper we prove this fact
for $r=0$ (Section \ref{bigger}) and $s=0$, $k=1$ (Sections \ref{smaller1},
\ref{smaller2}).

Let $d$ be the grading operator, $d(g\otimes t^n) = n g\otimes t^n$.
We can decompose the irreducible representation $L$ into the sum of
$d$-graded components $L(0)$, $L(1), \dots$. Denote by $D_jL$ the
sum $L(0) \oplus \dots \oplus L(j)$. We have a filtration $\{D_jL\}$
of $L$. For any subalgebra ${\frak b}$, consider the induced filtration of the
space $I(L,{\frak b})$. Namely, let $D_jI$ be the
image of $D_jL$ in $I(L,{\frak b})$. Let $c_j = \dim(D_jI/ D_{j-1}I)$.
We call the polynomial $\sum c_j q^j$ the $q$-dimension of
the space $I(L,{\frak b})$. In fact, using the bi--grading of $L$
(see Section 2) we can obtain
a polynomial in two variables. Such polynomials are certain ``finitizations" of
the character of the integrable representation $L$.

In this paper we partially answer the following questions for the special
cases described above (see \cite{FF}, \cite{FL}
for more about the same subject).

1) How to find explicitly the set of graded (or bi--graded) vectors in
$L$ such that their images form a basis in $I(L,{\frak b})$?

2) How to find explicit formulas for $q$-dimensions?
There are two types of such formulas: ``bosonic'' and ``fermionic''
 (\cite{KKMM}).
In this text we present only fermionic formulas for $k=1$.

3) How to find a $q$-version of Verlinde's rule which gives
a recursive description of the spaces of coinvariants?

Unfortunately, in general, it is not known how to investigate the filtration
on the space $I(L,{\frak b})$ explicitly. Let us describe our strategy
in the special situation considered in this paper.

Our subalgebra ${\frak b}$ is a member of the family of Lie subalgebras
depending on geometric data $(z_1,\ldots,z_s;w_1,\ldots,w_r)$, which constitute
a configuration space ${\cal C}$. We consider the limit to the special
point $(0,\ldots,0;0,\ldots,0)\in{\cal C}$. At this
point the subalgebra ${\frak b}^0$ and therefore the quotient
space $I(L,{\frak b}^0)$ is bi--graded. We find that the dimension does not
change: $\dim I(L,{\frak b}) = \dim I(L,{\frak b}^0)$. This implies that
the adjoint graded space ${\rm Gr}\,I(L,{\frak b})$ is canonically isomorphic
to $I(L,{\frak b}^0)$
as bi--graded space. In other words, we can, in principle, construct a basis
of ${\rm Gr}\,I(L,{\frak b})$ which remains valid as a basis in the limit.

In the case $r=0$, it is enough to work with the subalgebra in this limit
to obtain the answers to the above questions. In the case $s=0$ we need
to deform not only the subalgebra ${\frak b}$ in $\wsl$ but also $\wsl$ itself
in ${{\widehat{\mathfrak{sl}}_3}}$, as explained below.

Consider the Lie algebra $\mathfrak{sl}_3$ and its irreducible representation $\CC^3$.
Let us decompose the space $\CC^3$ into the sum of two subspaces $V_1$ and
$V_2$, where $\dim V_1 = 1$. Then the subalgebra $\mathfrak{sl}_2 \subset \mathfrak{sl}_3$ consists
of operators which annihilate $V_1$ and map $V_2$ to itself.

Consider the deformation of subspaces $V_1(\varepsilon)$, $V_2(\varepsilon)$,
$\varepsilon \in \CC$ such that $V_1(1) = V_1$, $V_2(1) = V_2$, for any
$\varepsilon \ne 0$ we have $V_1(\varepsilon) + V_2(\varepsilon) = \CC^3$
and for $\varepsilon=0$ we have $V_1(0) \subset V_2(0)$. Let
${\frak a}(\varepsilon)\subset \mathfrak{sl}_3(\varepsilon\not=0)$ be the subalgebra of
operators which annihilates $V_1(\varepsilon)$ and maps $V_2(\varepsilon)$ to
$V_2(\varepsilon)$. Clearly, ${\frak a}(\varepsilon)$ is isomorphic to $\mathfrak{sl}_2$
if $\varepsilon\ne0$, and
$$
{\frak a}(0){\buildrel{\rm def}\over=}
{\lim}_{\varepsilon\rightarrow0}{\frak a}(\varepsilon)
$$
is the maximal nilpotent subalgebra in  $\mathfrak{sl}_3$, so it is isomorphic to
the $3$-dimensional Heisenberg algebra.

Let $L$ be an integrable representation of
$\wsl\simeq\widehat{{\frak a}}(\varepsilon)$. We can deform it
inside an integrable representation of $\widehat{\mathfrak{sl}}_3$, and consider the limit
to a representation of $\widehat{{\frak a}}(0)$. It is known
(\cite{FS},\cite{St}) that the $q$-dimension of the limit $L(0)$
coincides with that of $L$.

We can also deform a graded subalgebra ${\frak b}^0 \subset \wsl$ and obtain
a deformed space of coinvariants $I(L(0),{\frak b}^0(0))$ in the limit
$\varepsilon\rightarrow0$. We will prove for the case $k=1$ that the dimension
of the space of coinvariants does not change in this limit.

In \cite{FS}  the dual space to $L(0)$ is realized in terms of symmetric
functions in two groups of variables. This functional description of
$L(0)^*$ was used in \cite{FS} to get a fermionic (Gordon) type 
formula for the character of $L$. In this paper we use the same approach
to coinvariants in the $k=1$ case. From this point of view we propose a
finitization of the results from \cite{FS}.

In this paper we avoid using $\mathfrak{sl}_3$ and present  an equivalent way to
obtain the Lie algebra
$\widehat{{\frak a}}(0)$ from $\wsl$. Namely, we introduce a filtration on
$\wsl$ such that the adjoint graded Lie algebra is $\widehat{{\frak a}}(0)$.

The plan of the paper is as follows. In Section 2 we formulate the problem.
We use a simpler definition of $\widehat{\mathfrak{sl}}_2$, i.e., using the ring
of Laurent polynomials. In Section 3 we give a monomial basis of the space
of coinvariants in the case $r=0$ with arbitrary level $k$. In Sections 4 and 5
we study the case $s=0$, and give monomial bases for the space of coinvariants
with $k=1$. In Appendix we prove Verlinde's rule in the form we use in this
paper.

{\bf Acknowledgements.}\quad
RK, TM, EM thank MSRI and the organizers of the Semester on
Random Matrices and Applications, and BF, SL, RK thank RIMS,
for a pleasant and stimulating environment.
BF thanks E. Frenkel for the hospitality at UC, Berkeley.
EM thanks E. Frenkel and A. Varchenko for useful discussions.
The authors thank M. Jimbo for careful reading of the manuscript
and giving a few suggestions for improvement of the text.
\section{Spaces of coinvariants}\label{definitions}
\subsection{Preliminary definitions}
Let $\slt=\mathfrak{sl}_2\otimes \C[t,t^{-1}] \oplus \C c \oplus \C d$ be the
affine $\mathfrak{sl}_2$ algebra with the commutator given by
\bea\notag
\lefteqn{[g_1\otimes p_1(t,t^{-1}),g_2\otimes p_2(t,t^{-1})]=}\\
&&
[g_1,g_2]\otimes p_1p_2-c\,(g_1,g_2)\,\res_{t=0}(\frac{dp_1}{dt}p_2)
=[g_1,g_2]\otimes
p_1p_2+c\,(g_1,g_2)\,\res_{t=\infty}(\frac{dp_1}{dt}p_2).\notag 
\ena
where $g_1,g_2\in \mathfrak{sl}_2$, and $(g_1,g_2)$ stands for the Killing form
computed on $g_1,g_2$. Here $c$ stands for the
central element and $d$ the degree operator, $[d,x_i]=i x_i$.
We note that $\res_{t=\infty}t^i=-\delta_{i,-1}$ in our convention.

Let $\{e,f,h\}$ be the standard basis of $\mathfrak{sl}_2$,
$$
e=\pmatrix{0&1\cr0&0},\quad
f=\pmatrix{0&0\cr1&0},\quad
h=\pmatrix{1&0\cr0&-1}.
$$ 
Then the algebra $\slt$ has a basis
$\{x_i=x\otimes t^i,c,d \ |\ x=e,f,h\in \mathfrak{sl}_2, i\in \Z\}$.
We will also use the generating current notation, $e(z)=\sum_{i\in\Z}
e_i z^i$, etc..  Note that
$e_i,f_i,h_i$ ($i=0,1$) are {\it not} the Chevalley generators.

{\it Remark 1.}\quad
Let $s=t^{-1}$. Then we have another presentation of the same algebra
(which is standard in the literatures),
$\slt=\mathfrak{sl}_2\otimes \C[s,s^{-1}] \oplus \C c \oplus \C d$ and the
commutator takes the form
\be
[g_1\otimes p_1(s,s^{-1}),g_2\otimes p_2(s,s^{-1})]=
[g_1,g_2]\otimes p_1p_2+c\,(g_1,g_2)\,\res_{s=0}(\frac{dp_1}{ds}p_2),
\ee

{\it Remark 2.}\quad
Note that in the main text, we employ a purely algebraic approach,
and therefore use the definition of
$\widehat{\mathfrak{sl}}_2$ with the Laurent polynomial ring without completion.
This is different from Introduction and Appendix, where the geometric
aspect is vital and we need the completion.

\medskip
Consider the Verma module $\verm=U(\slt) v_{k,l}$, with $k\in\Z_{>0}$
and $0\leq l \leq k$, generated by the highest weight vector
$v_{k,l}$ satisfying
\begin{eqnarray}
&&cv_{k,l}=kv_{k,l},\quad h_0v_{k,l}=lv_{k,l},\quad dv_{k,l}=0,\nonumber\\
&&e_iv_{k,l}=0\ (i\leq 0),\quad h_iv_{k,l}=f_iv_{k,l}=0\ (i<0).\label{sing}
\end{eqnarray}
Note that there is a well-defined action of $e(z)^{k+1}$ on $M_{k,l}$,
since the infinite sum expressions appearing in the coefficients of this
series in $z$ reduce to finite sums while acting on any vector.

The irreducible highest weight module
$\inte$ is the quotient of $\verm$ by the 
the submodule generated by the singular vectors $e_1^{k-l+1}
v_{k,l}$ and $f_0^{l+1} v_{k,l}$. The module $\inte$ is called the
integrable module of level $k$ and weight $l$.

\begin{proposition}(\cite{FS}, \cite{FM}) The module $\inte$ is the quotient of the module
$\verm$ by the subspace generated by the vectors
\bean
e(z)^{k+1}w \quad ({\rm or\;\; equivalently,\;\,} f(z)^{k+1}w)
\quad{\rm where}\quad w\in\verm.
\eean
\end{proposition}
\begin{proof} By \cite{FM}, the series $e(z)^{k+1}$ and  $f(z)^{k+1}$
act on $L_{k,l}$
by $0$. It follows that $e_1^{k-l+1}v_{k,l}=0$. Indeed, we have
$e(z)^{k+1}v_{k,l}=0$ and by \Ref{sing} we obtain
$e_1^{k+1}v_{k,l}=0$. Applying $f_{-1}$ to this equation $l$ times and
using \Ref{sing} again we obtain $e_1^{k-l+1}v_{k,l}=0$.

Similarly, from $f(z)^{k+1}=0$, it follows that $f_0^{l+1}v_{k,l}=0$. Note
that $[f_0,e(z)]=-h(z)$, $[f_0,h(z)]=2f(z)$ and $[f_0,f(z)]=0$, so 
$f(z)^{k+1}=0$ , because a scalar multiple of $f(z)^{k+1}$ is obtained
from $e(z)^{k+1}$ by 
successive commutators with $f_0$.\qed
\end{proof}

Given ${\cal Z}=(z_1,\ldots,z_N)\in\C^N$, $z_i$ not necessarily distinct,
we define two types of subalgebras of $\mathfrak{sl}_2\otimes\C[t]$
and their corresponding spaces of coinvariants.
If the points are distinct,
the subalgebras are special cases of subalgebras defined in Introduction.
In general, we define the subalgebras corresponding to points
in completed moduli spaces (though we do not discuss completion in this paper).

The first subalgebra is
\bean
\mathfrak{sl}_2({\cal Z};0)&=&\mathfrak{sl}_2\otimes(t-z_1)\cdots(t-z_N)\C[t],
\eean
which reduces to $0$ at each point $t=z_i$.
(Here the symbol $0$ means the trivial Lie subalgebra of $\mathfrak{sl}_2$.)
The corresponding space of coinvariants is defined by
\bea
L^{(N)}_{k,l}({\cal Z};0)&=&L_{k,l}/\mathfrak{sl}_2({\cal Z};0)L_{k,l}.
\label{FIRST}
\ena

The second subalgebra is
\bean
\mathfrak{sl}_2({\cal Z};{\frak n})&=&
\C e\otimes\C[t]\oplus(\C h\oplus\C f)\otimes
(t-z_1)\cdots(t-z_N)\C[t].
\eean
This subalgebra reduces to ${\frak n}=\C e\subset \mathfrak{sl}_2$ at $t=z_i$.
The corresponding space of coinvariants is defined by
\bea\label{SECOND}
L^{(N)}_{k,l}({\cal Z};{\frak n})&=&L_{k,l}/\mathfrak{sl}_2({\cal Z};{\frak n})L_{k,l}.
\ena

In each case, the spaces of coinvariants are finite-dimensional, and their
dimensions are given in terms of the Verlinde algebra (see Appendix).
Let us recall its definition.

Let $\pi_l$ be the irreducible $(l+1)$-dimensional representation
of $\mathfrak{sl}_2$.
The level $k$ Verlinde algebra associated to $\mathfrak{sl}_2$, which we denote by $\Vk$,
is an algebra over $\C$, with generators $\{\pi_l;\ 0\leq l \leq k\}$
and multiplication
\begin{equation}\label{VERLINDE}
\pi_l\cdot\pi_{l'}=\sum_{i=|l-l'|\atop i+l-l':\hbox{\rm even}}
^{\hbox{\rm min}(2k-l-l',l+l')}\pi_i.
\end{equation}
Define the positive integers $d^{(N)}_{k,l}(0)$ and $d^{(N)}_{k,l}({\frak n})$ by
\bean
(\pi_0+2\pi_1+\cdots+(k+1)\pi_k)^N&=&\sum_ld^{(N)}_{k,l}(0)\pi_l,\\
(\pi_0+\pi_1+\cdots+\pi_k)^N&=&\sum_ld^{(N)}_{k,l}({\frak n})\pi_l.
\eean
We have
\setcounter{subsection}{1}
\begin{theorem}\label{Ver}
If the points $z_1,\ldots,z_N$ are distinct, then
\bean
\hbox{\rm dim}\,L^{(N)}_{k,l}({\cal Z};0)&=&d^{(N)}_{k,l}(0),\\
\hbox{\rm dim}\,L^{(N)}_{k,l}({\cal Z};{\frak n})&=&d^{(N)}_{k,l}({\frak n}).
\eean
\end{theorem}

Theorem \ref{Ver} is proved in Appendix.

\begin{example}
When $k=1$,
\bean
d_{1,l}(0)=\frac{3^N+(-1)^{N+l}}2,\qquad
d_{1,l}({\frak n})=2^{N-1}, \quad l=0,1.
\eean
\end{example}

We also define a special case of the coinvariants \Ref{FIRST},
\Ref{SECOND}, at ${\cal Z}=(0,\ldots,0)$:
\bean
L^{(N)}_{k,l}(0)\buildrel{{\rm def}}\over=
L^{(N)}_{k,l}/\mathfrak{sl}_2(0)L^{(N)}_{k,l},\\
L^{(N)}_{k,l}({\frak n})\buildrel{{\rm def}}\over=
L^{(N)}_{k,l}/\mathfrak{sl}_2({\frak n})L^{(N)}_{k,l}.
\eean
where
\bean
\mathfrak{sl}_2(0)=\hbox{\rm lim}_{z_1,\ldots,z_N\rightarrow0}
\mathfrak{sl}_2({\cal Z};0),\\
\mathfrak{sl}_2({\frak n})=\hbox{\rm lim}_{z_1,\ldots,z_N\rightarrow0}
\mathfrak{sl}_2({\cal Z};{\frak n}).
\eean

These spaces inherit a bi-grading with respect to
$(h_0,d)$ from $L_{k,l}$.  We define the character of any of these 
bi-graded space $V$ by
\bean
\hbox{\rm ch}_{q,z}V&=&\hbox{\rm trace}_V\left(q^dz^{h_0}\right).
\eean

\def \coi {L_{k,l}^{(N)}({\cal Z};0)} 
\def\frakB{{\hat{\frak n}_+}}
\section{The space of coinvariants $\coi$.}\label{bigger}
\def \mod {{\rm mod}}

In this section we consider the space of coinvariants $\coi$ 
(see (\ref{FIRST}))
where $z_1, \dots, z_N$  are not necessarily distinct points in ${\Bbb C}$. 
We introduce a monomial basis for this space, and derive a recursion relation
for the characters and dimensions of these coinvariants. We prove that
at the special point ${\cal Z}=0$, the dimension of the coinvariants is
equal to the Verlinde number as in the case of generic ${\cal Z}$.

\subsection{Monomial basis}

Here we propose a monomial basis of $\coi$, and of the whole
representation $L_{k,l}$.

Both the space of coinvariants $\coi$ and the $\widehat{\mathfrak{sl}}_2$-module
 $L_{k,l}$ are quotients of the Verma module 
$M_{k,l}= U(\frakB) v_{k,l}$, where $\frakB = \mathfrak{sl}_2\otimes
t{\Bbb C}[t] \oplus f_0$ is the nilpotent part of the Borel subalgebra of
$\widehat{\mathfrak{sl}_2}$. 
Define an ordering of the basis of $\frakB$ in the following way:
$$ f_0 < e_1 < h_1 < f_1 < e_2 < h_2 < f_2 < \cdots.$$

\begin{definition}
A monomial $m$ in $U(\frakB)$ is called {\it ordered}
if it is of the form
\begin{equation} \label{MONO}
f_{n-1}^{a_{n-1}} h_{n-1}^{b_{n-1}} e_{n-1}^{c_{n-1}} f_{n-2}^{a_{n-2}} 
\cdots 
e_2^{c_2} f_1^{a_1} h_1^{b_1} e_1^{c_1} f_0^{a_0}.
\end{equation}
\end{definition}

By the PBW Theorem, the ordered monomials form a basis of $U(\frakB)$.

In the following, we will not make a notational distinction
between the highest weight vector $v_{k,l} \in M_{k,l}$
and its image in its quotients, e.g., $L_{k,l}$.
By abuse of language, we will also call a vector 
of the form $m\cdot v_{k,l}$ $(m\hbox{ a monomial in }U(\frakB))$ a monomial.

\begin{definition}
Let $k$ and $l$ be fixed integers. 
We call an ordered monomial of the form \Ref{MONO} an {\it $ehf$--monomial} 
if it satisfies the following conditions:
\def \theenumi {\roman{enumi}}
\def\labelenumi {(\theenumi)}

\begin{enumerate}
\item $a_i+a_{i+1}+b_{i+1}\leq k$ for all $i\geq 0$,
\item $a_i+b_{i+1}+c_{i+1}\leq k$ for all $i\geq 0$,
\item $a_i+b_i +c_{i+1}\leq k$ for all $i>0$, 
\item $b_i + c_i + c_{i+1} \leq k$ for all $i>0$,
\item $a_0 \leq l$, $c_1 \leq k-l$.
\end{enumerate}

\end{definition}

\begin{remark}
Conditions (i)--(iv) can be expressed by the following picture:

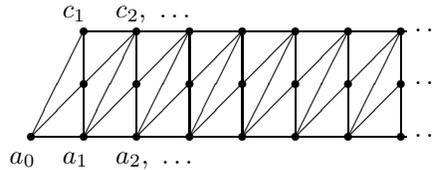
\begin{center}
\begin{picture}(180,70)

\multiput(40,20)(0,40){2}{\line(1,0){122}}
\put(20,20){\line(1,0){20}}
\multiput(40,20)(20,0){7}{\line(0,1){40}}
\multiput(20,20)(20,0){7}{\line(1,1){20}}
\multiput(20,20)(20,0){7}{\line(1,2){20}}
\multiput(40,40)(20,0){6}{\line(1,1){20}}
\multiput(165,20)(0,20){3}{\dots}
\multiput(20,20)(20,0){8}{\circle*{3}}
\multiput(40,40)(20,0){7}{\circle*{3}}
\multiput(40,60)(20,0){7}{\circle*{3}}
\put(12,10){$a_0$}
\put(32,10){$a_1$} 
\put(52,10){$a_2,\ \dots$}
\put(32,65){$c_1$}
\put(52,65){$c_2, \ \dots$}

\end{picture}
\end{center}              

We consider monomials 
$$f_{n-1}^{a_{n-1}} \cdots h_1^{b_1} e_1^{c_1} f_0^{a_0},$$
such that the sum of exponents over any
triangle (corresponding to the conditions (i)--(iv))
is less than or equal to $k$.
\end{remark}

\begin{definition}
An $ehf$--monomial has {\it length $N$} if
$a_i = b_i = c_i = 0$ for $i \geq N$.
\end{definition}

Our goal is to prove that the set of $ehf$--monomials forms a basis in
$L_{k,l}$, and that the set of $ehf$--monomials of length $N$ forms a basis in
$\coi$.

Define a linear isomorphism from the polynomial ring
$R=\C[f_0,e_1,h_1,f_1,\ldots]$ to $U(\frakB)$,
$m\mapsto:m:$, such that the inverse image of an ordered monomial
$m\in U(\frakB)$ is $m\in R$. Namely, for a given element in $R$,
we specify by (\ref{MONO}) how to order the non-commuting product
in $U(\frakB)$. Note that we not only order $x_i$ ($x=e,f,h$) with respect to
the index $i$ but also put $e_i,h_i,f_i$ from the left to the right
in this order. 
The existence of such isomorphism is a consequence of
the PBW theorem.

We often abuse the notation $m$ to mean both an ordered monomial
of $U(\frakB)$ and the corresponding element in $R$.

Next, we define a complete lexicographic ordering
on the set of ordered monomials. (Note the difference between
``lexicographic ordering'' and ``ordered monomials.'')
We define the homogeneous degree ${\rm Deg}\,m$ of an ordered monomial $m$:
\bean
{\rm Deg}\,m&=&\sum_i(a_i+b_i+c_i).\label{Deg}\\
\eean

If ${\rm Deg}\,m>{\rm Deg}\,m'$, we say that $m>m'$.
If ${\rm Deg}\,m={\rm Deg}\,m'$, we compare the exponents $a_0$, $c_1$,
$b_1$, $a_1$, etc., in this order.
Namely, we say $m>m'$ if the value of $a_0$
in $m$ is {\it smaller} than that in $m'$; if these values are equal,
then we compare $c_1$ next, and so on.

We define the dominant monomial of an element $u\in U(\frakB)$
to be the largest monomial in the expression of $u$
written as a linear combination of ordered monomials. We note that
if $m_i$ $(i=1,2)$ is the dominant monomial of $u_i$, then the 
dominant monomial of the product $u_1u_2$ is $:m_1m_2:$.

We will now consider the list of elements in $U(\frakB)$ which are
trivial in $L_{k,l}=U(\frakB)v_{k,l}$.
We will show that the dominant monomial of each element in this list
breaks one or more of the conditions (i)--(v).
The remaining monomials, which are not dominant monomials of
these trivial elements, span the space $L_{k,l}$.

\begin{lemma}\label{rel}
Let
\bean
\psi_{r,s} = ({\rm ad}\,e_0)^s\sum_{{i_1, \dots, i_{k+1} \geq
0}\atop{i_1+\dots+i_{k+1} = r}} f_{i_1} f_{i_2} \dots
f_{i_{k+1}}\in U(\frakB).
\eean
Then $\psi_{r,s}v_{k,l}=0$ in $L_{k,l}$.
\end{lemma}

\begin{proof}
Since $f(z)^{k+1}=0$ in $L_{k,l}$, all
$d$--graded elements of the form
$$\psi_r=\sum_{i_1+ \dots +i_{k+1}=r} f_{i_1} f_{i_2} \dots
f_{i_{k+1}}$$
act trivially in $L_{k,l}$. When acting on
the highest weight vector $v_{k,l}$,
$$\psi_{r}v_{k,l} =  \sum_{{i_1, \dots, i_{k+1} \geq
0}\atop{i_1+\dots +i_{k+1} = r}} f_{i_1} f_{i_2} \dots
f_{i_{k+1}} v_{k,l}=0\in L_{k,l}.$$
In addition, since $e_0v_{k,l}=0$,
$$\psi_{r,s}v_{k,l}= ({\rm ad}\,e_0)^s(\psi_r)v_{k,l}=0.\qed$$
\end{proof}

It is easy to find the dominant monomial of $\psi_{r,s}$. For example,
the dominant monomial of $\psi_{r,0} = \psi_r$ is $f_{\alpha+1}^\beta
f_\alpha^{k+1-\beta}$, where $r = \alpha(k+1)+\beta$ and $0\leq \beta \leq
k$.

\begin{theorem}\label{span} 
The set of $ehf$--monomials spans $L_{k,l}$.
\end{theorem}

\begin{proof}

Consider the set of elements in $U(\frakB)$,
\begin{equation}
m\cdot\psi_{r,s}\quad(2k+2 \geq s \geq 0, \ r\geq s/2)
\label{LIST1}
\end{equation}
and
\begin{equation}
m\cdot f_0^{l+1},m\cdot e_1^{k-l+1},
\label{LIST2}
\end{equation}
where $m$ is an ordered monomial.

Due to Lemma \ref{rel} and the condition \Ref{sing}, 
when applied to the highest weight vector $v_{k,l}$,
these are trivial in $L_{k,l}$. 
Let us show that an ordered monomial is
the dominant monomial of an element in this set if and only if
it violates one or more of the conditions (i)--(v).

The ordered monomials $:m\cdot f_0^{l+1}:$ and
$:m\cdot e_1^{k-l+1}:$ are the dominant monomials of 
$m\cdot f_0^{l+1}$ and $m\cdot e_1^{k-l+1}$, respectively, and
an ordered monomial violates the condition (v) if and only if it can
be represented in one of these forms.

Consider $\psi_{r,s}$ with $r=\alpha(k+1)+\beta$ and $0\leq\beta\leq
k$. Their dominant monomials violate one of the conditions (i)--(iv),
as shown in the following table:

\begin{center}
\begin{tabular}{|c|c|c|}
\hline
\rule{0mm}{5mm}
Case & Dominant monomial of $\psi_{r,s}$ & Violates condition \\ 
\hline
\rule{0mm}{5mm}
$s \leq \beta$ & 
$f_{\alpha+1}^{\beta-s}
h_{\alpha+1}^sf_\alpha^{k+1-\beta}$ & (i)\\
\hline
\rule{0mm}{5mm}

$\beta \leq s \leq 2\beta$ &
$h_{\alpha+1}^{2\beta-s}
e_{\alpha+1}^{s-\beta}f_\alpha^{k+1-\beta}$ & (ii)\\
\hline
\rule{0mm}{5mm}

$2\beta \leq s \leq k+1+\beta$ & 
$e_{\alpha+1}^{\beta}f_\alpha^{k+1+\beta-s}
h_{\alpha}^{s-2\beta}$ & (iii)\\
\hline
\rule{0mm}{5mm}

$k+1+\beta \leq s$ &
$e_{\alpha+1}^{\beta}h_\alpha^{2k+2-s}
e_\alpha^{s-(k+1)-\beta}$ & (iv)\\
\hline
\end{tabular} 
\end{center}
Given any ordered monomial $m$, the dominant term in
$m\cdot\psi_{r,s}$ also violates the conditions as listed above, hence
it is not an $ehf$-monomial.

On the other hand, suppose that an ordered monomial $m$ violates one of the
conditions (i)--(iv). It can decomposed in $R$ as $m=m_1\cdot m_2$,
in such a way that ${\rm Deg}\, :m_2:=k+1$, and
$:m_2:$ violates the same condition as $m$.
Then, using this table, there exist $r$ and $s$ such that $m_2$ is the
dominant monomial of $\psi_{r,s}$. Since $m$ is the dominant monomial of
$:m_1:\psi_{r,s}$, any monomial which violates any of the conditions
(i)--(iv) is of the form $m\cdot\psi_{r,s}$.

We have shown that the set of $ehf$-monomials, which satisfy
conditions (i)--(v) span the space $L_{k,l}$.\qed
\end{proof}

\begin{theorem}\label{span2}
The set of $ehf$--monomials of length $N$ spans $\coi$.
\end{theorem}

\begin{proof} The proof is similar to Theorem \ref{span}.
Recall that $\coi$ is a quotient of $L_{k,l}$, and therefore of
$M_{k,l}=U(\frakB)v_{k,l}$.
For any $g \in \mathfrak{sl}_2$, $i\geq 0$, define $g^{(i)}=
g\otimes t^i(t-z_1)\dots(t-z_N) \in U(\frakB)$. 
By definition it is trivial in $\coi$. If $g = e$, $h$ or $f$, then the
dominant monomial of $g^{(i)}$ is $g_{i+N}$.

Consider the set of elements
$$\{e^{(i)}\cdot m,h^{(i)}\cdot m,f^{(i)}\cdot m\} \in U(\frakB),$$
together with the elements (\ref{LIST1}), (\ref{LIST2}).
Clearly, their image in $\coi$ is zero. 
The dominant monomials of $g^{(i)}\cdot m$ $(g=e,h,f;i\geq 0)$ are precisely
the monomials of length larger than $N$. Therefore, our assertion follows.
\qed
\end{proof}

\subsection{Dimension of the coinvariants}

We have shown that the set of $ehf$--monomials of length $N$ spans the
space of coinvariants $\coi$.  In order to prove that they are indeed
linearly independent, we compare the dimension of $\coi$ with the
number of all $ehf$--monomials of length $N$.

If ${\cal Z} = (z_1, \dots, z_N)$  are all distinct complex numbers,
the dimension of $\coi$ is given by the Verlinde numbers $d_{k,l}^{(N)}(0)$
(see Theorem \ref{Ver}). These satisfy a recursion relation.

\begin{lemma}\label{count}
The Verlinde numbers $d_{k,l}^{(N)}(0)$ satisfy the recursion relation
\bea
d_{k,l}^{(N+1)}(0)&=&\sum_{l'} {\cal D}^{l'}_l d_{k,l'}^{(N)}(0),\quad
N\geq 1;\label{RECD}\\
d_{k,l}^{(1)}(0)&=&l+1,\label{INIVEC}
\ena
where
\bea
{\cal D}^{l'}_l =
(\min(l, k-l')+1) \cdot (\min(l', k-l)+1).
\label{RECUR}
\ena
\end{lemma}

\begin{proof}
The equation (\ref{RECD}) follows from (\ref{VERLINDE}) and
the fact that ${\cal D}^i_j$ is the sum of the dimensions of the
representations appearing in the right hand side of this formula.
\qed
\end{proof}

In other words, we have the matrix ${\cal D} = \left( {\cal D}^{l'}_l
\right)$ and vectors ${\bold d}_k^{(N)} = \left(d_{k,l}^{(N)}(0)\right)$
such that ${\bold d}_k^{(N+1)} = {\cal D}\cdot {\bold d}_k^{(N)}$, or,
equivalently, 
${\bold d}_k^{(N)} = {\cal D}^{N-1} \cdot {\bold d}_k^{(1)}$. 

Let $p(k,l,N)$ be the number of $ehf$--monomials of length $N$. In order
to prove that these monomials are linearly independent, we must show that
$d_{k,l}^{(N)}(0) = p(k,l,N)$. 
The proof is based on the following technical lemma for reduction of the size
of matrices.

\begin{lemma}\label{red}
Let $A=(A^\alpha_{\alpha'})$ be an $n\times n$ matrix, and
$v$, $\phi$ vectors of length $n$. Suppose that for some
indices $\alpha_0, \alpha_1, \dots, \alpha_s$, 
$\phi_{\alpha_i}= \phi_{\alpha_j}$ and 
$A^{\alpha}_{\alpha_i} = A^{\alpha}_{\alpha_j}$ for
any $i,j,\alpha$. Then for any $m$ we have
$$ \sum_{i,j=1,\ldots,n} \phi_i \left( A^m \right) ^i_j v^j =
\sum_{i,j=1,\ldots,n-s} 
\tilde{\phi}_i \left(\tilde{A}^m\right)^i_j \tilde{v}^j,$$  
where $\tilde{\phi}$ is obtained from $\phi$ by deleting
$\phi_{\alpha_i}$ for $i\not=0$, $\tilde{v}$ from $v$ by deleting
$v^{\alpha_i}$ for $i\not=0$ and replacing $v^{\alpha_0}$ with
$\sum_{i\geq 0} v^{\alpha_i}$, and $\tilde{A}$ from $A$ by deleting
the rows and columns with indices $\alpha_i$ for $i\not=0$ and replacing
$A^{\alpha_0}_{\alpha'}$ with $\sum_{i\geq 0} A^{\alpha_i}_{\alpha'}$.
\end{lemma}

Sometimes we will use this lemma for the case
$\phi_i=\tilde\phi_i=1$ for all $i$. In this case we omit writing $\phi_i$
or $\tilde\phi_i$.

\begin{proposition}
\label{mcount}
$d_{k,l}^{(N)}(0) = p(k,l,N)$. 
\end{proposition}

\begin{proof}
Let $p^{a,b,c}(k,l,N)$ be the number of $ehf$--monomials of length
$N$ with $a_{N-1}=a$, $b_{N-1}=b$, $c_{N-1}=c$. 
Conditions (i)--(v) imply that 
\begin{equation}\label{abc}
p^{a,b,c}(k,l,N+1) = \sum_{{a',b',c'}\atop
{{a'+a+b\leq k,\ a'+b+c\leq k}\atop
{a'+b' +c\leq k,\  b' + c' + c \leq k}}} p^{a',b',c'}(k,l,N).
\end{equation}
\def \vp {{\bold p}}
Consider the matrix ${\cal P}=({\cal P}^{a,b,c}_{a',b',c'})$ given by
\bean
{\cal P}^{a,b,c}_{a',b',c'} =
\cases{
1&if\quad$a'+a+b\leq k$, $a'+b+c\leq k$, $a'+b'+c\leq k$ and $b'+c'+c\leq k$;
\cr
0&otherwise.}
\eean
Note that this matrix does not depend on $l$.
Clearly,
$p(k,l,N) = \sum_{a,b,c}p^{a,b,c}(k,l,N)$. 

{}From (\ref{abc}) we have
\begin{equation}\label{e1}
p(k,l,N) = \sum_{{a,b,c}\atop{a',b',c'}}
\left({\cal P}^{N-1} \right)^{a,b,c}_{a',b',c'}p^{a',b',c'}(k,l,1),\qquad N>1,
\end{equation}
with
$$
p^{a,b,c}(k,l,1) =
\cases{
1&if\quad$c=0$ and $a+b=l$;\cr
0&otherwise.}
$$

Let us simplify this formula using Lemma \ref{red}. 
Note that ${\cal P}^{a,b,c}_{a',b',c'}$ depends only on $a$, $b$, $c$,
$a'$ and $m'=\max(a'+b', b'+c')$.
So we can apply Lemma \ref{red}
to all the multi--indices $a, b, c$ with the same $a$ and $m$
in (\ref{e1}).
We obtain
\begin{equation}\label{e2}
p(k,l,N) =  \sum_{{0\leq a\leq m\leq k}\atop{0\leq a' \leq m' \leq k}} 
\left( \widetilde{\cal
P}^{N-1}\right)^{a,m}_{a',m'}\widetilde{p}^{a',m'}(k,l,1),
\end{equation}
\bean
\widetilde{\cal P}^{a,m}_{a',m'} &= &
\cases{
\quad 0&if\quad$a'+m>k$;\cr
\min(k-m',m)+1&otherwise,}
\\
\widetilde{p}^{a',m'} &=&
\cases{
1&if\quad$m'=l$;\cr
0&otherwise.}
\eean

For example, if $k=1$, we have
\bean
{\cal P}=
\pmatrix{
1&1&1&1&1\cr
1&0&0&0&0\cr
1&1&1&0&0\cr
1&1&1&0&0\cr
1&0&0&0&0}
;\quad
\tilde{\cal P} =
\pmatrix{
1&1&1\cr
2&1&0\cr
2&1&0}
;\quad
D=
\pmatrix{
1&2\cr
2&1}
.
\eean

See also (\ref{MAT}) (with $q=z=1$) and the matrix $\tilde{\cal P}$
in the proof.

Note that $\widetilde{\cal P}^{a,m}_{a',m'}$ 
and $\widetilde{p}^{a,m}$ do not depend on $a$,
so we can transpose (\ref{e2}) and then apply 
Lemma \ref{red} again. The
reduced matrix is in fact equal to the matrix ${\cal D}$ of
(\ref{RECUR}), and the reduced column is nothing but (\ref{INIVEC}):
$$p(k,l,N) = \sum_{0 \leq m' \leq k}\left( {\cal D}^{N-1}
\right)^l_{m'}\cdot(m'+1)=d_{k,l}^{(N)}(0).\qed$$
\end{proof}

\begin{theorem}\label{ttt}
{\ }
\def \theenumi {\roman{enumi}}
\def\labelenumi {(\theenumi)}
\begin{enumerate}
\item  The set of $ehf$--monomials forms a basis of $L_{k,l}$.
\item  The set of $efh$--monomials of length $N$ forms a basis of $\coi$
for an arbitrary set of points ${\cal Z}$.
\end{enumerate}
\end{theorem}

\begin{proof}
Let us begin with (ii).  If ${\cal Z}$ consists of distinct complex numbers
then (ii) follows from Theorem \ref{span2} and Proposition \ref{mcount}.
Arguments from deformation theory show that for any (not necessarily distinct)
${\cal Z}$ we have $\dim L^{(N)}_{k,l}(0) \geq \dim \coi\geq d^{(N)}_{k,l}(0)$. 
On the other hand, Theorem \ref{span2} for ${\cal Z} = (0,\dots,0)$ shows that 
$\dim L^{(N)}_{k,l}(0) \leq p(k,l,N)$. Therefore, by using 
Proposition \ref{mcount}, we have the equality of dimensions for any
${\cal Z}$. The statement (ii) follows.

The statement (i) follows from Theorem \ref{span} and (ii); we need only to
prove that $ehf$--monomials are linearly independent. Suppose that some
linear combination of $m_1,\ldots,m_s$ is trivial in $L_{k,l}$.
Let $N$ be the maximal length of monomials $m_i$. Then, this linear combination
is trivial in $\coi$, which contradicts (ii).\qed
\end{proof}

\subsection{Recursion relations for characters of the coinvariants}

As an application of Theorem \ref{ttt} we can write a recursion relation
for ${\rm ch}_{q,z}L^{(N)}_{k,l}(0)$:

\def \ch {{\rm ch}}

\begin{corollary}
$$\ch_{q,z} L^{(N)}_{k,l}(0) = \sum_{0\leq a \leq l} z^{l-2a}
\left( {\cal P}^{(N)}\right)^{0,0}_{a,l}(q,z),$$
where 
$$ {\cal P}^{(N)} (q,z) = \tilde{\cal P}(q^N,z)\cdot\tilde{\cal P}(q^{N-1},z)
\cdots\tilde{\cal P}(q^2,z)\cdot\tilde{\cal P}(q,z),$$
and the matrix $\tilde{\cal P}$ is given as follows:
$$
\tilde{\cal P}^{a,m}_{a',m'}(q,z) =0\quad\hbox{if $a'+m>k$;}
$$
otherwise,
\begin{equation}\label{mat}
\tilde{\cal P}^{a,m}_{a',m'}(q,z) =
\cases{
q^m z^{-2a}\cdot\frac{1-(qz^2)^{k-m'+1}}{1-qz^2}&if\quad$k-m'\leq a$;\cr
q^m z^{-2a}\left(\frac{1-(qz^2)^a}{1-qz^2}+(qz^2)^a\cdot
\frac{1-z^{2(k-m'-a+1)}}{1-z^2}\right)
&if\quad$a\leq k-m'\leq m$;\cr
q^m z^{-2a} \left( \frac{1-(qz^2)^a}{1-qz^2} + (qz^2)^a\cdot
\frac{1-z^{2(m-a+1)}}{1-z^2}\right)
&if\quad$m\leq k-m'$.}
\end{equation}
\end{corollary}

\begin{proof}
Each monomial (\ref{MONO}) contributes $z^{2\alpha}q^\beta$ to the character
where $\alpha = \sum_i (c_i-a_i)$, $\beta = \sum_ii\cdot(a_i+b_i+c_i)$;
there is also a contribution from $v_{k,l}$, which is equal to $z^l$.

Let $p^{a,b,c}(k,l,N;q,z)$ be the partial character
of the linear span of $ehf$--monomials of length $N$ 
with $a_{N-1}=a$, $b_{N-1}=b$, $c_{N-1}=c$. 
Then we have
$$p^{a,b,c}(k,l,N+1;q,z) = 
\sum_{{a',b',c'}\atop
{{a'+a+b\leq k,\ a'+b+c\leq k}\atop
{a'+b' +c\leq k,\  b' + c' + c \leq k}}}
q^{N(a+b+c)}z^{2(c-a)}p^{a',b',c'}(k,l,N;q,z),$$
and therefore
\def \cp {{\cal P}}
$$ \ch_{q,z} L^N_{k,l}(0) =   \sum_{{a,b,c}\atop {a',b',c'}}
\left( \cp(q^{N-1},z)\cdot \cp(q^{N-2},z)\cdots\cp(q,z)
\right)
^{a,b,c}_{a',b',c'} 
\cdot p^{a',b',c'}(k,l,1;q,z), $$
where
\bea
&&{\cal P}^{a,b,c}_{a',b',c'}(q,z)\nonumber\\
&& =\cases{
q^{(a+b+c)}z^{2(c-a)}&if\quad$a'+a+b\leq k$, $a'+b+c\leq k$, $a'+b'+c\leq k$
and $b'+c'+c\leq k$;\cr
0&otherwise.}\nonumber
\ena

The coefficients ${\cal P}^{a,b,c}_{a',b',c'}(q,z)$ 
depend only on $a$, $b$, $c$, $a'$ and $m' = \max(a'+b', b'+c')$.
Therefore, we can apply Lemma \ref{red} again. We obtain
\def \tp {{\tilde{\cal P}}}
$$ \ch_{q,z} L^N_{k,l} (0) =   \sum_{{0\leq a \leq m}\atop {0 \leq a' \leq
m'}}
\left( \tp(q^{N-1},z)\cdot \tp(q^{N-2},z)\cdots\tp(q,z)
\right)^{a,m}_{a',m'}\cdot\tilde{p}^{a',m'}(k,l,1;q,z), $$
with
$$
\tilde{p}^{a,m}(k,l,1;q,z) =
\cases{
z^{l-2a}&if\quad$m=l$;\cr
0&otherwise.}
.
$$
Finally, noting that $\tilde{\cal P}^{0,0}_{a,m} (q,z)= 1$ for all $a$ and $m$,
we can replace the summation with respect to $a$ and $m$ with the
multiplication of the row vector $\tilde{\cal P}^{(0,0)}(q^N,z)$.\qed
\end{proof}

When $k=1$, the matrix $\tilde{\cal P}$ is
\bea
\pmatrix{
1&1&1\cr
q(1+z^2)&q&0\cr
qz^{-2}+q^2&qz^{-2}&0}
.
\label{MAT}
\ena

Let us write the characters for $k=1$. The proof will be supplied elsewhere.
Let $D$ be the matrix
$$
D = \pmatrix{
2 &1 & 0\cr
1 & 2 & 1 \cr
0 & 1 & 2}
$$
\def \be {{\bold e}}
\def \bn {{\bold n}}
and $\be\in\Z^3$. Define
$$c(N,\be) = \sum_{{\bn=(n_1,n_2,n_3)}\atop{n_1, n_2, n_3 \geq 0}} 
 z^{2(n_3-n_1)} q^{\frac12\bn D \bn + \be\bn} \prod_{i=1}^3 
\qbin[N+1-(D\bn - n + \be)_i; n_i].$$
Then
\bean
\left( {\cal P}^{(N)}\right)^{0,0}_{0,0}(q,z) = c(N,(0,0,0)),\\
\left( {\cal P}^{(N)}\right)^{0,0}_{0,1}(q,z) = c(N,(0,0,1)),\\
\left( {\cal P}^{(N)}\right)^{0,0}_{1,1}(q,z) = c(N,(1,1,1)).
\eean
The characters of the coinvariant $L^N_{k,l}(0)$ in the case $k=1$ are
\bean
\ch_{q,z} L_{1,0}^{(N)}(0) &=& c(N,(0,0,0)), \\
\ch_{q,z} L_{1,1}^{(N)}(0) &=& z\cdot c(N,(0,0,1)) +z^{-1}\cdot c(N, (1,1,1)).
\eean

\section{The space of coinvariants $L^{(N)}_{k,l}({\frak n})$:
Monomial basis and dual space}\label{smaller1}

\def \hh {{\cal H}}
\def \adj {{\rm Gr}}

\subsection{Filtration and Heisenberg modules.}
First, introduce the filtration $\{F_j\wsl\}$ on the
Lie algebra $\wsl$ where $F_{-1} \wsl =0$, $F_0\wsl=\C c\oplus\C d$,
$F_1\wsl=F_0\wsl+\sum_{i\in\Z}\C e_i+\sum_{i\in\Z}\C f_i$ and $F_2\wsl = \wsl$.
Then we consider the adjoint graded Lie algebra
$\adj\wsl=\C c\oplus \C d\oplus \oplus\hh$ where $\hh$ has the basis
$\{ \bar e_i,\bar h_i,\bar f_i \}$ such that 
\begin{equation}\label{comm}
[\bar e_i,\bar f_j]=\bar h_{i+j} ,\qquad
[\bar e_i,\bar h_j]=[\bar f_i,\bar h_j]=
[\bar f_i,\bar f_j]=[\bar e_i,\bar e_j]=0.
\end{equation}
Therefore, $\hh$ is the algebra of currents
with values in the 3--dimensional Heisenberg Lie algebra.

This filtration induces a filtration on $U(\wsl)$. Clearly, we have
$\adj U(\wsl) \cong U(\adj \wsl)$.

Now consider the representation $L_{1,0}$ of $\wsl$. Let $v$ be its
highest vector, then $L_{1,0} = U(\wsl) v$. We have the filtration
$F_i L_{1,0} = (F_i U(\wsl)) v$. The algebra $\hh$ acts on the adjoint
graded space $\adj L_{1,0}$. Since in $L_{1,0}$ we have $e^2(z)=0$ and
$f^2(z)=0$, in $\adj L_{1,0}$ we have $\bar e^2(z)=0$,
$\bar f^2(z)=0$ and due to (\ref{comm}) we have $\bar e(z) \bar h(z)=0$,
$\bar f(z) \bar h(z) = 0$, $\bar h^2(z) = 0$.

Let $M$ be the representation of $\hh$ induced from the trivial
1--dimensional representation $\CC \bar v$ of the subalgebra $\hh^-$
spanned by $\bar e_i$, $\bar h_i$, $\bar f_i$ with $i\leq 0$. Let
$W$ be the quotient of $M$ by the subspace generated by the elements
of the form
\begin{equation}\label{relations}
\bar e^2(z)m, \quad \bar f^2(z)m, \quad \bar e(z) \bar h(z)m,
\quad \bar f(z) \bar h(z)m, \quad \bar h^2(z)m\quad(m\in M).
\end{equation}
Clearly, $\adj L_{1,0}$ is a quotient of $W$ and below we prove that $\adj
L_{1,0} \cong W$.

\def \ZZ {{\mathbb Z}}

The algebra $\hh$ and the space $W$ admit a $\ZZ^3$--grading. Namely, we
set
\bean
\deg_{z_1} \bar e_i = 1,& \deg_{z_1} \bar h_i = 1,& \deg_{z_1} \bar f_i =0
\\
\deg_{z_2} \bar e_i = 0,& \deg_{z_1} \bar h_i = 1,& \deg_{z_1} \bar f_i =1
\\
\deg_{q} \bar e_i = i,& \deg_{q} \bar h_i = i,& \deg_{q} \bar f_i =i.
\eean

Then we have the grading on $M$ and, as relations~(\ref{relations}) are
homogeneous, on $W$. Denote the corresponding grading operators by $H_1$,
$H_2$ and $d$ respectively.
We define the character of the space $W$ by
\bean
{\rm ch}_{q,z_1,z_2}W&=&{\rm trace}q^dz_1^{H_1}z_2^{H_2}
\eean
and similarly for any graded subspace or quotient of $W$.
Let $W_{m,n}$ be the graded component of degree $(m,n)$ with respect to
$(H_1,H_2)$.

\begin{theorem}\label{monomialbasis}
The following elements span
$W_{m,n}$:
\begin{equation}\label{basis}
v_{\alpha,\beta}=\prod_{j=1}^{m}\bar e_{\alpha_j+2(m-j)+1-n}
\prod_{j=1}^n \bar f_{\beta_j+2(n-j)+1}\ \bar  v,
\end{equation}
where $\alpha,\beta$ are partitions (i.e., $\alpha_1\geq \alpha_2\geq \cdots
\geq \alpha_m\geq 0$).
\end{theorem}

First we prove a technical lemma.
We use the notation $F_i^{(n)}$ for a sum of terms of the
form $\bar f_{i_1}\cdots \bar f_{i_n}$, with degree $\sum i_j=i$.
\begin{lemma}\label{reorder}
$$
\bar e_{-n-a} F_i^{(n)}\bar v = \sum_{b\geq 0} \bar e_{1-n+b}
F_{i-b-a-1}^{(n)}\bar v\hbox{ for $a\geq 0$}.
$$
\end{lemma}
\begin{proof}
Consider the case $n=1$. We have
$$\bar e_{-1-a} \bar f_i \bar v = [\bar e_{-1-a}, 
\bar f_i ]\bar v = [\bar e_0,\bar f_{i-a-1}]\bar v
= \bar e_0 \bar f_{i-a-1}\bar v.
$$
Then use induction on $n$. Consider
\begin{eqnarray*}
\bar e_{-n-a} \bar f_b F_i^{(n-1)}\bar v &=& 
(\bar f_b \bar e_{-n-a} + [\bar e_{1-n}, \bar
f_{b-1-a}])F_i^{(n-1)}\bar v \nonumber \\
& &\hbox{\hskip-1in} =
 \bar f_b\sum_{c\geq 0} \bar e_{2-n+c} F_{i-2-c-a}^{(n-1)} \bar v+
\bar f_{b-1-a}\sum_{c\geq 0} \bar e_{2-n+c}F_{i-c-1}^{(n-1)}\bar v+
\bar e_{1-n}F^{(n)}_{b+i-1-a}\bar v  
.\label{step}
\end{eqnarray*}
In the second line, we assumed the lemma is proved for $n-1$. 
The last term is of the desired form, and the other terms can be reduced
to this form by using the following procedure:
\begin{eqnarray*}
\bar f_a \bar e_{2-n+b} F_c^{(n-1)}\bar v &=& ( \bar e_{2-n+b} \bar
f_a+[\bar f_{a+1+b},\bar e_{1-n}]) F_c^{(n-1)}\bar v \\
&&\hbox{\hskip-1in}=\bar f_{a+1+b}\sum_{d\geq 0}  \bar e_{2-n+d}
F^{(n-1)}_{c-1-d}\bar v + 
(\bar e_{2-n+b}\bar f_a - \bar e_{1-n}\bar f_{a+1+b}) F_c^{(n-1)}\bar v.
\end{eqnarray*}
In the second step, the last two terms are again of the desired form,
and the degree of $F^{(n-1)}_{c-d-1}$ in the first term is less than $c$, the
degree on the left hand side of the equation. Therefore if we repeat
this step a finite number of times, the process will terminate when the
degree is low enough so that $F^{(n-1)}_{c-d-1}\bar v=0$.
We are left with a sum of terms with $\bar e_i$
acting on the left, with degree $i\geq  1-n$. Therefore the lemma
follows.\qed
\end{proof}

\noindent{\it Proof of Theorem \ref{monomialbasis}.}
Clearly, $W_{m,n}$ is spanned by elements of the form 
\begin{equation}
\bar e_{i_1}\cdots \bar e_{i_{m-l}}
\bar h_{j_1}\cdots \bar h_{j_l} \bar f_{k_1}\cdots \bar f_{k_{n-l}} \bar v,
\label{spanning}
\end{equation}
where all the indices are positive. 

Also, any element of the form \Ref{spanning} can be expressed as a sum
of terms of the form
\begin{equation}
\label{one}
\bar e_{i_1}\cdots \bar e_{i_m}\bar f_{j_1}\cdots \bar f_{j_n}\bar v,
\hbox{ with $j_a>0$},
\end{equation}
rewriting $\bar h_i = [\bar e_{-a},\bar f_{i+a}]$ with $a$
sufficiently large, and using the fact
that $\bar h_i$ commute with all other generators.

The relations $\bar e(z)^2=\bar f(z)^2=0$ in $W$ mean that \Ref{one} can be
rewritten as a sum of similar elements, with indices which satisfy
\begin{equation}\label{sep}
|i_{k}-i_{k+1}|\geq 2\quad{\rm and}\quad|j_{k}-j_{k+1}|\geq 2.
\end{equation}

Now, using Lemma \ref{reorder}, any element of the form
$\bar e_{a}\prod_{i=1}^n \bar f_{j_i} \bar v$ with $a<1-n$ can be  
rewritten as a sum
of such terms with $a\geq  1-n$.
The result is a sum of terms of the form
$$
\prod_{k=1}^m \bar e_{i_k}\prod_{k=1}^n \bar f_{j_k}\bar v,\quad i_\alpha\geq  1-n.
$$
We again use $\bar e(z)^2=\bar f(z)^2=0$ so that the indices again
satisfy \Ref{sep}. Note that this step preserves the degree with respect to
the grading operator $d$ of $\prod \bar e_j$ and $\prod \bar f_j$, separately,
whereas using the lemma, the degree of the product of $\bar e$'s is raised
and $\bar f$'s lowered. Therefore, repeating the process a finite number of
times, we obtain a sum of terms of the form \Ref{basis}.
$\;\qed$

We next show that in fact, \Ref{basis} form a basis for
$W_{m,n}$. In order to show linear independence, we introduce the dual space.

\subsection{Dual space}

Define the space $S_{m,n}$ to be the space of rational functions in
two sets of variables, ${\bf x}=(x_1,\ldots,x_m)$ and
${\bf y}=(y_1,\ldots,y_n)$, of the form
\begin{equation}\label{F}
F({\bf x},{\bf y})=
\prod_{i=1}^m x_i \prod_{i=1}^n y_i
\frac{\prod_{i<j}(x_i-x_j)^2
\prod_{i<j}(y_i-y_j)^2}
{\prod_{i,j}(x_i-y_j)}f({\bf x},{\bf y}),
\end{equation}
where $f({\bf x},{\bf y})$ is a symmetric polynomial in the variables
${\bf x}$ and ${\bf y}$ separately. 

Consider the variables $\{z_1,\ldots,z_{m-l}\},
\{u_1,\ldots,u_{l}\}, \{w_1,\ldots,w_{n-l}\}$ in some order $\sigma$. 
Let $\iota: ({\bf x,y})\to
({\bf z, u, w})$ be defined as follows: Arrange the letters $z_i,u_j$ 
in the order in which they appear in $\sigma$. Then $\iota({\bf x})$ is
this sequence. Similarly, ${\bf y}$ is mapped to the ordering
in $\sigma$ of $w_i,u_j$. For example, if the sequence of variables is 
$(u_1 w_1 z_1 u_2)$, then $\iota({\bf x})=(u_1,z_1,u_2)$ and
$\iota({\bf y})=(u_1,w_1,u_2)$.

We define a coupling of an element in $S_{m,n}$ to an element in
$W_{m,n}$ by using generating functions for elements in $W_{m,n}$,
\begin{eqnarray}\label{coupling}
\left< F({\bf x,y}) , \sigma\left(\prod_{i=1}^{m-l} \bar e(z_i)
\prod_{i=1}^{l} \bar h(u_i) \prod_{i=1}^{n-l}
\bar f(w_i)\right)\bar v\right>& =&\nonumber\\ & & {\hskip-2in}
\ \left. \left(\prod_{\{i,j;\ \iota(x_i)=\iota(y_j)\}}
x_i^{-1}(x_i-y_j) \right)
F({\bf x},{\bf y})\right|_{{\bf x}\mapsto\iota({\bf x})
\atop{\bf y}\mapsto\iota({\bf y})},
\end{eqnarray}
where $\sigma(p_1\cdot\dots\cdot p_{n+m-l})$ means the product of
elements $p_i$ in the order $\sigma$.

Namely, we first remove the pole in (\ref{F})
at $x_i=y_j$, for $x_i$ and $y_j$ that are mapped by $\iota$ to the same
variable of type $u$, then we make the substitution.
After this substitution, the remaining pole factors are only of the form
$(z_i-w_j)^{-1}$. They are understood as follows.
If the variable $z_i$ appears to the right of the variable $w_j$,
then we expand the rational function in positive powers of $z_i$.
We expand in negative powers if it appears to the left of $w_j$. We
will use the notation 
$$
\frac{1}{w-z}= w^{-1}\sum_{n\geq 0} (z/w)^n,\quad 
\frac{1}{z-w}= z^{-1}\sum_{n\geq 0} (w/z)^n.
$$
Note that
\bea
\frac{zw}{z-w}+\frac{zw}{w-z}&=&z\delta(z/w),
\label{delta}
\ena
where $\delta(z)=\sum_{i\in\Z}z^i$ is the multiplicative delta function.

For example, the coupling \Ref{coupling} means
\begin{eqnarray}
\left< F(x_1,y_1),\bar e(z)\bar f(w)\bar v\right> &=& \frac{z w}{z-w}f(
z,w),\nonumber\\ 
\left< F( x_1,y_1),\bar f(w)\bar e(z)\bar v\right> &=& 
-\frac{z w}{w-z}f(z,w),\nonumber \\
\left< F( x_1;y_1),\bar h(u)\bar v\right> &=& uf(u;u).
\label{example}
\end{eqnarray}

\begin{lemma}
The coupling \Ref{coupling} is well defined.
\end{lemma}

\begin{proof} 
The fact that $\langle F,g_i\bar v\rangle=0$ $(g=e,f,h\in \mathfrak{sl}_2;i\leq 0)$
follows from the fact that 
the right hand side of \Ref{coupling} contains only positive powers of
the rightmost variable in the product of currents acting on $\bar v$.

The fact that $F({\bf x,y})$ is a symmetric function in $x_i$ and $y_i$
implies that 
$$\langle F,\cdots[\bar e(z),\bar e(w)]\cdots \bar v\rangle=
\langle F,\cdots[\bar f(z),\bar f(w)]\cdots \bar v\rangle=0.$$
The same argument holds for $\langle
F,[\bar e(z),\bar h(w)]\bar v\rangle= \langle F,[\bar f(z),\bar h(w)]\bar v\rangle=0$.

Since $F({\bf x,y})=0$ if $x_1=x_2$ or $y_1=y_2$, we have 
$$\langle F,\cdots e(z)^2\cdots 
\bar v\rangle=\langle F,\cdots \bar f(z)^2\cdots \bar v\rangle = 0.$$
>From \Ref{delta} and \Ref{example} it is clear that
$\langle F,\cdots([\bar e(z),\bar f(w)]-\delta(z/w)\bar h(z))\cdots
\bar v\rangle=0$.

Therefore, the lemma follows.\qed
\end{proof}

We define a complete (lexicographic) ordering among all partitions, 
\begin{equation}\label{order}
\lambda > \mu \ {\rm if}\ \lambda_i=\mu_i,\ i=1,\ldots,j-1 \ {\rm and}\
\lambda_j > \mu_j
\end{equation}
for some $j$. 
We extend the ordering to pairs of partitions $\{(\alpha,\beta)\}$
by setting $(\alpha,\beta)>(\alpha',\beta')$ if $\alpha>\alpha'$, or
$\alpha=\alpha'$ and $\beta>\beta'$.

We pick the basis for the symmetric polynomials $f({\bf x,y})$ to be
$$ 
f_{\alpha,\beta}({\bf x,y}) = ({\bf x}^\alpha {\bf y}^\beta)^{\rm Symm},
$$
where the symmetrization is with respect to ${\bf x}$ and ${\bf y}$
separately, and we use the standard notation
$$
{\bf x}^\alpha {\buildrel{\rm def}\over=}x_1^{\alpha_1}\cdots x_m^{\alpha_m}.
$$
Note that the separately symmetric polynomials
$f_{\alpha,\beta}$ are ordered by the ordering given above.
We extend this ordering even for the separately symmetric Laurent polynomials
since we will use it in that way in the proof of Lemma \ref{triangular}.

We define $F_{\alpha,\beta}({\bf x,y})$ to be a function
of the form \Ref{F} with $f$ replaced by $ f_{\alpha,\beta}$. These functions
form a basis of $S_{m,n}$.

\begin{lemma}\label{triangular}
The coupling \Ref{coupling} between $F_{\alpha,\beta}$ and $v_{\alpha,\beta}$
of \Ref{basis} is triangular, with nonzero diagonal elements.
\end{lemma}
\begin{proof}
We define an ordering among all separately symmetric Laurent polynomials
of the form
$\left(
{\bf x}^{\bf \alpha}
{\bf y}^{\bf \beta}\right)^{\rm Symm}$
in a similar manner as for pairs
of partitions $({\bf \alpha},{\bf \beta})$.
Then, we have
$$
F_{\alpha,\beta}({\bf x,y})=\left(\prod_{i=1}^m x_i^{\alpha_i+2(m-i)+1-n}
\prod_{j=1}^n y_j^{\beta_j+2(n-j)+1}\right)^{\rm Symm} + \ {\rm lower
\ terms}.
$$
The coupling \Ref{coupling} is equivalent to
\begin{equation}\label{coup2}
\langle ({\bf x}^{\bf i'}
{\bf y}^{\bf j'})^{\rm Symm},
\bar e_{i_1}\cdots \bar e_{i_m}\bar f_{j_1}\cdots \bar f_{j_n}\bar v\rangle = 
\delta_{\bf i,i'}\delta_{\bf j,j'}.
\end{equation}
Therefore the coupling of the basis $F_{\alpha,\beta}$ to the vectors
$v_{\alpha,\beta}$ is triangular.\qed
\end{proof}

\begin{theorem}\label{DUAL}
The coupling \Ref{coupling} is nondegenerate, and
the space $S_{m,n}$ is dual to $W_{m,n}$.
\end{theorem}
\begin{proof}
This follows from Theorem \ref{monomialbasis} and Lemma \ref{triangular}.\qed
\end{proof}

\begin{corollary}\label{BASEW}
The vectors $\{v_{\alpha,\beta}\ ; \alpha,\beta \ {\rm partitions}\}$ of
\Ref{basis} form a basis of $W$.
\end{corollary}

Set
$$
S_{m,n}(d)=\{F({\bf x},{\bf y})\in S_{m,n};
F(t{\bf x},t{\bf y})=t^dF({\bf x},{\bf y})\}.
$$
Using Theorem \ref{DUAL}, we have
$$
{\rm ch}_{q,z_1,z_2}W = \sum_{m,n,d} 
{\rm dim}S_{m,n}(d) q^d z_1^m z_2^n= 
\sum_{m,n\geq 0}\frac{q^{n^2+m^2-mn} }{(q)_m (q)_n}z_1^m z_2^n.
$$
When specialized to $z_1=z_2^{-1}=z^2$, this is the character of $L_{1,0}$.
After summation, we obtain
$$
{\rm ch}_{q,z}L_{1,0}=\sum_{n\in\Z}\frac{q^{n^2}z^{2n}}{(q)_\infty}.
$$
as noted in [FS]. 
And as $\adj L_{1,0}$ is a quotient of $W$ and $\ch\, \adj L_{1,0} = \ch
L_{1,0}$ we have 

\begin{corollary}
$W \cong \adj L_{1,0}$ as graded $\hh$--modules.
\end{corollary}

\subsection{The dual to the coinvariants of $W$}
Let ${\frak g}(M,N)\subset \hh\otimes\C[t,t^{-1}]$
be the subalgebra
generated by $\{\bar e_M,\bar e_{M+1},\ldots,\bar f_N,\bar f_{N+1},\ldots\}$,
and define
\begin{equation}\label{coinv}
\overline{W}^{M,N}= W/{\frak g}(M,N)W\quad
\hbox{for $M,N\geq -1,(M,N)\not=(-1,-1)$}.
\end{equation}
We further define $W^{M,N}=\overline{W}^{M,N}$ if $M,N\geq  0$, and
$W^{M,N}= \overline{W}^{M,N}/\C\bar v$ if $M=-1$ or $N=-1$.
The quotient $W^{M,N}$ inherits the same $\ZZ^3$--grading as $W$ itself.

\begin{lemma}
The orthogonal complement to the space
${\frak g}(M,N)W$ in $S_{m,n}$ is the subspace $\overline S^{M,N}_{m,n}$, 
consisting of $F$ (see \Ref{F}) with the degree restrictions
\begin{equation}\label{degrees1}
{\rm deg}_{x_1} F({\bf x,y})<M, \qquad 
{\rm deg}_{y_1} F({\bf x,y})<N.
\end{equation}
\end{lemma}
If $m=0$ there is no restriction on
${\rm deg}_{x_1} F({\bf x,y})$, and the same is true for $n$.
The degree of $F$ is counted in such a way that
\bea
{\rm deg}_x{1\over x-y}={\rm deg}_y{1\over x-y}=-1.
\ena
\begin{proof}
We only note that we expand
\bea
{1\over z-w}={1\over z}\sum_{n=0}^\infty\left({w\over z}\right)^n
\ena
when we compute the coupling of $F$ with an element of the form
$\bar e(z)\cdots\bar f(w)\cdots\bar v$.\qed
\end{proof}

In terms of $f$ (see \Ref{F}) the restrictions read as
\begin{equation}\label{degrees}
{\rm deg}_{x_1} f({\bf x,y}) \leq  M-2m+n, \qquad 
{\rm deg}_{y_1} f({\bf x,y}) \leq  N-2n+m.
\end{equation}
Note that $\overline S^{M,N}_{0,0}=\C$ for all $M,N$.

We define $S^{M,N}_{m,n}$ to be the dual space of $W^{M,N}_{m,n}$.
For $M,N\geq 0$, we have $S^{M,N}_{m,n}=\overline S^{M,N}_{m,n}$.
For $M=-1$ or $N=-1$, $S^{M,N}_{m,n}$ is the subspace of
$\overline S^{M,N}_{m,n}$, which is codimension $1$ and $S^{M,N}_{0,0}=0$.

The reason for the above definition of the space $W^{M,N}_{m,n}$
and its dual at $(m,n)=(0,0)$ will be clear in Section \ref{smaller2}.

We have
\begin{theorem}\label{coinbasis}
The following vectors form a basis for $W^{M,N}_{m,n}$ for $(m,n)\not=(0,0)$:
\begin{equation}\label{basis1}
\bar e_{i_1}\cdots \bar e_{i_m} \bar f_{j_1}\cdots \bar f_{j_n}\bar v
\end{equation}
with $i_k\geq  i_{k+1}+2$, $j_k\geq  j_{k+1}+2$, $M>i_k\geq  1-n$ and
$N+m> j_k >0$.
\end{theorem}
This space is finite dimensional and the character is given by
\begin{eqnarray}
{\rm ch}_{q,z_1,z_2}W^{M,N}&=&
\sum_{m,n\geq 0}q^{n^2+m^2-mn}\qbin[M-m+n;m]\qbin[N-n+m;n]z_1^mz_2^n.
\label{coinchar} 
\end{eqnarray}
\begin{proof}
Since we have Corollary \ref{BASEW}, it is enough to show, first that
a vector of the form (\ref{basis1})
belongs to the orthogonal complement to the space $\bar S^{M,N}_{m,n}$
if $i_1\geq M$ or $j_1\geq N+m$,
and second that the characters of the space of vectors of the form
(\ref{basis1}) satisfying the conditions in the statement of the theorem,
is equal to that of the space $S^{M,N}_{m,n}$. We only note that $N$
is replaced by $N+m$ because $\bar f_{j_1}$ is placed to the right of
$\bar e_{i_1}\cdots\bar e_{i_m}$. The rest of the proof is straightforward.\qed
\end{proof}

\subsection{Dimension of the space $L^{(N)}_{1,0}({\frak n})$}

Here we prove that the space of coinvariants $L^{(N)}_{1,0}({\frak n})$  
is isomorphic to the space of coinvariants $W^{0,N} \cong \adj
L_{1,0}/{\frak g}(0,N)$ as bi--graded spaces. To do it we need

\begin{lemma}\label{deformation}
Let $V$ be a vector space with filtration
$$
0=F_{-1}V\subset F_0 V \subset F_1 V \subset\cdots\subset V, \qquad
V=\bigcup_j
F_j V.   
$$  
Let $T_i:V\to V$, $(i\in I)$, be a set of linear maps with degree
$d_i\geq0$,
i.e. $T_i(F_jV)\subset F_{j+d_i}V$.  

Let $\adj V=\oplus_j\;{\rm Gr}_j V =\oplus_j\; F_j V /F_{j-1} V$ and
let  $\overline{T}_i$ be the induced graded maps. Then, we have the
inequality for the dimensions
of the spaces of coinvariants:
\bea\label{deformation inequality}
{\rm dim}\;\left( V/\sum_iT_iV \right) \leq
{\rm dim}\; \left(\adj V /\sum_i\overline{T}_i\,
\adj V\right).
\ena
\end{lemma}

\begin{proof}
Choose a set of vectors $\{v^n_j\in F_n V;j\in J_n\}$ such that their 
images form
a basis in
the space $\adj V/\sum_i\overline{T}_i \adj V$.
We prove by induction on $n$ that the images of the elements
$\{v^s_j; s\leq n, j\in J_s\}$ span $F_n V/\sum_iT_iF_{n-d_i} V$. Indeed,
let
$v\in F_{n+1} V$. Then there exists a linear combination $\sum
a_jv^{n+1}_j$,
$a_j\in\C$, such that the image of $v-\sum a_jv^{n+1}_j$ in
${\rm Gr}_{n+1} V$ is in the image $\sum_i\overline{T}_i \adj V$.
It means that $v-\sum a_jv^{n+1}_j=u+w$, where $u\in\sum_i T_iV$
and $w$ is in $F_n V$. By induction hypothesis there exists a linear
combination
 $\sum b_jv^s_j$, $b_j\in\C$, $s=s(j)\leq n$, such that
$w-\sum b_jv^s_j\in\sum_i T_iV$. Then,
$v-\sum a_jv^{n+1}_j-\sum b_jv^s_j\in\sum_i T_iV$.

Therefore, if the images of $\{v^n_j\in V, n\in\Z, j\in J_n\}$ form
a basis in $\adj V/\sum_i\overline{T}_i \adj V$ then the
images of the same vectors span $V/\sum_iT_iV$.\qed
\end{proof}

\begin{remark}
Note that if by some other argument inequality \Ref{deformation
inequality} is actually an equality, and  
the images of $\{v_j\in V,\;\;j\in J\}$ form a basis in
$\adj V/\sum_i\overline{T}_i \adj V$, 
then the images of the same vectors form a basis in $V/\sum_iT_iV$.
\end{remark}
 
\begin{corollary}\label{cc1}
$\dim L^{(N)}_{1,0}({\frak n}) \leq \dim W^{0,N}$
\end{corollary}

\begin{proof}
It follows from Lemma
\ref{deformation} applied to 
the vector space $L_{1,0}$ with the filtration defined in section 4.1.1,
and the set of operators $e_i,\, i\geq 0$; $f_i,\, i\geq N$.\qed
\end{proof}

\begin{corollary}\label{cc2}
$\dim L^{(N)}_{1,0}({\cal Z}, {\frak n}) \leq \dim L^{(N)}_{1,0}({\frak
n})$
\end{corollary}

\begin{proof}
It follows from Lemma \ref{deformation} applied
to 
the vector space $L_{1,0}$ with the filtration induced by the $d$-grading,
and the set of operators
$$
\{e\otimes\ t^i, f\otimes\prod_{j=1}^N(t-z_j)t^{i},
\;i\geq 0\},
$$
where ${\cal Z} = (z_1, \dots, z_N)$.\qed
\end{proof}

\begin{theorem}\label{charform}
The character of the coinvariants $L^{(N)}_{1,0}({\frak n})$ is given by
\bea\label{ch}
{\rm ch}_{q,z}L^{(N)}_{1,0}({\frak n})&=&
\sum_{s\geq 0}q^{s^2} \qbin[N;2s]z^{-2s}.\label{hil}
\ena
We have, in particular,
\bean
{\rm dim}\,L^{(N)}_{1,0}({\frak n})&=&2^{N-1}.
\eean
\end{theorem}
\begin{proof}
By expanding the identity $(a+b)^N=(a+b)^{N-s}(a+b)^s$ with $ba=q ab$,
we have
\bean
{\rm ch}_{q,z_1,z_2}W^{0,N}|_{z_1=z_2^{-1}=z^2}
&=&\sum_{n\geq  s\geq 0}q^{s^2+n(n-s)}\qbin[N-s;n]\qbin[s;n-s]z^{-2s}\\
&=&\sum_{s\geq 0}q^{s^2}\qbin[N;2s]z^{-2s}.
\eean
We have, in particular, that
\bean
{\rm dim}W^{0,N}&=&2^{N-1}.
\eean
This is equal to the dimensions of the coinvariants
$L^{(N)}_{1,0}({\cal Z};{\frak n})$ given by Verlinde's formula.

So by Corollary~\ref{cc1} and Corollary~\ref{cc2} we have that $\dim
L^{(N)}_{1,0}({\frak n}) = \dim W^{0,N}$. Note that we have the same
inequalities as in Lemma~\ref{deformation} for any graded component of
these spaces. Therefore we have the equality of characters, that is the
statement of the theorem.\qed
\end{proof}

\begin{corollary}
The vectors of the form
\bean
e_{i_1}\cdots e_{i_m} f_{j_1}\cdots f_{j_n}\bar v_{1,0}
\eean
with the same conditions as in Theorem \ref{coinbasis} form a monomial basis
for the space of coinvariants $L_{1,0}^{(N)}({\frak n})$. 
\end{corollary}

\section{Recursion relations for the spaces $W^{M,N}$ and the dual spaces}
\label{smaller2}

\subsection{Short exact sequences}
In this section we describe the spaces of coinvariants $W^{0,N}$ using
induction on $N$. This will allow us to produce another monomial basis
for the space $L^{(N)}_{1,0}({\frak n})$.
We also write down the recurrence relations
for the corresponding characters. To establish the induction we use the
spaces $W^{M,N}$ with $M$ general.

Introduce the maps
\bean
T^*&:&S^{M,N}_{m,n}\to S^{M,N+1}_{m,n},\\
U^*&:&S^{M,N}_{m,n}\to S^{M+1,N}_{m,n},\\
\phi^*&:&S^{M,N}_{m,n}\to S^{M+1,N-2}_{m,n-1},\\
\epsilon^*&:&S^{M,N}_{m,n}\to S^{M-2,N+1}_{m-1,n}
\eean
as follows. For $F({\bf x},{\bf y})\in S^{M,N}_{m,n}$, define
\bean
T^*(F({\bf x},{\bf y}))&=&y_1y_2\dots y_n F({\bf x},{\bf y}),\\
U^*(F({\bf x},{\bf y}))&=&x_1x_2\dots x_m F({\bf x},{\bf y}),\\
\phi^*(F({\bf x},{\bf y}))&=&\left.\frac{1}{y_1}\;\frac{x_1x_2
\dots x_m}{y_2^2y_3^2\dots y_n^2}F({\bf x},{\bf y})\right|_{y_1=0},\\
\epsilon^*(F({\bf x},{\bf y}))&=&\left.\frac{1}{x_1}\;\frac{y_1y_2
\dots y_n}{x_2^2x_3^2\dots x_m^2}F({\bf x},{\bf y})\right|_{x_1=0}.
\eean

Recall that the spaces $S^{M,N}_{m,n}$ are defined only for $M,N\geq 
-1$, with $(M,N)\neq(-1,-1)$, and $m,n\geq  0$. We define
$S^{M,N}_{m,n}=0$ if $m<0$ or $n<0$. The maps
$T^*,U^*,\phi^*,\epsilon^*$ are defined only when both source and
target spaces are defined.

\begin{lemma}\label{short exact dual sequences} 
The maps $T^*,U^*,\phi^*,\epsilon^*$ are well defined. Moreover, the sequences
\bea\label{phi dual sequence 1}
0 \to S^{M,N-1}_{m,n} @>T^*>>S^{M,N}_{m,n} @>\phi^*>>S^{M+1,N-2}_{m,n-1} \to 
0,\qquad M\geq  -1,N\geq  1
\ena
and
\bea\label{phi dual sequence 2}
0 \to S^{M-1,N}_{m,n} @>U^*>>S^{M,N}_{m,n} @>\epsilon^*>>S^{M-2,N+1}_{m-1,n} 
\to 0, \qquad M\geq  1,N\geq  -1
\ena
are exact.
\end{lemma}
\begin{proof}
By symmetry reason it is enough to show one 
of these, say (\ref{phi dual sequence 2}).

It is clear that $U^*$ is well-defined and injective.

If $m\geq  1$, the map $\varepsilon^*$ sends $F$ of (\ref{F}) to
\bean
\varepsilon^*(F)&=&\prod_{i=2}^mx_i\prod_{i=1}^ny_i
\frac{\prod_{2\leq  i<j}(x_i-x_j)^2\prod_{i<j}(y_i-y_j)^2}
{\prod_{i\geq 2,j}(x_i-y_j)}f(0,x_2,\ldots,x_m;{\bf y}).
\eean
If $m\geq 2$, it is obvious that this is well-defined and surjective.
If $m=1$, it is less obvious because in the space $S^{M-2,N+1}_{0,n}$,
the first condition of (\ref{degrees1}) is void and also because
$S^{-1,N}_{0,0}\not=\overline S^{-1,N}_{0,0}$.

The first condition of (\ref{F}) for $S^{M,N}_{1,n}$ becomes
\bean
{\rm deg}_{x_1}x_1\prod_{i=1}^ny_i\frac{\prod(y_i-y_j)^2}
{\prod_{i=1}^n(x_1-y_i)}f(x_1;{\bf y})<M.
\eean
This is equivalent to ${\rm deg}_{x_1}f(x_1;{\bf y})<M+n-1$.
Therefore, if $n\geq 1$ or $M\geq 2$, then
\bean
{\rm deg}_{x_1}f(x_1;{\bf y})&=&0
\eean
is allowed and the map $\varepsilon^*$ is surjective.

If $n=0$ and $M=1$, we have $S^{1,N}_{1,0}=0$, and therefore,
we must use the definition $S^{-1,N+1}_{0,0}=0$.

The rest of the proof is straightforward.\qed
\end{proof}

Next, define the maps
\bean
T&:&W^{M,N}_{m,n}\to W^{M,N-1}_{m,n},\\
U&:&W^{M,N}_{m,n}\to W^{M-1,N}_{m,n},\\
\phi&:&W^{M,N}_{m,n}\to W^{M-1,N+2}_{m,n+1},\\
\epsilon&:&W^{M,N}_{m,n}\to W^{M+2,N-1}_{m+1,n}
\eean
as follows. Let $A=\sigma(\bar{e}_{i_1}\dots
\bar{e}_{i_a}\bar{h}_{s_1}\dots \bar{h}_{s_b}\bar{f}_{j_1}\dots
\bar{f}_{j_c})\bar{v}\in W^{M,N}_{m,n}$, where $\sigma$ denotes the
order in which $\bar{e}_i,\bar{h}_i$ and $\bar{f}_i$ are written, with
$a+b=m,\;\; b+c=n$. Define
\bean
T(A)&=&\sigma(\bar{e}_{i_1}\dots \bar{e}_{i_a}\bar{h}_{s_1-1}\dots
\bar{h}_{s_b-1}\bar{f}_{j_1-1}\dots \bar{f}_{j_c-1})\bar{v},\\
U(A)&=&\sigma(\bar{e}_{i_1-1}\dots \bar{e}_{i_a-1}\bar{h}_{s_1-1}\dots
\bar{h}_{s_b-1}\bar{f}_{j_1}\dots \bar{f}_{j_c})\bar{v},\\
\phi(A)&=&\sigma(\bar{e}_{i_1-1}\dots \bar{e}_{i_a-1}\bar{h}_{s_1+1}\dots 
\bar{h}_{s_b+1}\bar{f}_{j_1+2}\dots \bar{f}_{j_c+2})\bar{f}_1\bar v,\\
\epsilon(A)&=&\sigma(\bar e_{i_1+2}\dots \bar e_{i_a+2}\bar h_{s_1+1}\dots
 \bar h_{s_b+1}\bar f_{j_1-1}\dots \bar f_{j_c-1})\bar e_1\bar v.\\
\eean

\begin{lemma}
The maps $T,U,\phi,\epsilon$ are well defined.
\end{lemma}
\begin{proof}
Consider, for example, the map $\phi$. Clearly, $\phi$ preserves the
relations \Ref{comm},\Ref{relations} and the highest weight condition, $\bar
g_i\bar v=0$, where $g=e,f,h$ and $i\leq 0$.  For example,
the images of the zero vectors $\bar e_0\bar v$, $\bar f_0\bar v$, 
\bean
\phi(\bar e_0\bar v)=\bar e_{-1}\bar f_1\bar v, \qquad \phi(\bar f_0\bar v)=
\bar f_{2}\bar f_1\bar v,
\eean
are also zero vectors.

Finally, $\phi$ maps the subalgebra generated by $\{\bar
e_i,\bar f_j, i\geq  M,j\geq  N\}$ to the the subalgebra
generated by $\{\bar e_i,\bar f_j, i\geq  M-1,j\geq  N+2\}$.\qed
\end{proof}

We have another easy 
\begin{lemma}\label{dual maps}
The maps $T,U,\phi,\epsilon$ are dual to maps
$T^*,U^*,\phi^*,\epsilon^*$ with respect to pairing defined by \Ref{coupling}.
\end{lemma}

\begin{corollary}\label{short exact sequences}
The sequences
\bea\label{phi sequence}
0 \to W^{M+1,N-2}_{m,n-1} @>\phi>> W^{M,N}_{m,n} @>T>> W^{M,N-1}_{m,n}
\to 0,\qquad M\geq  -1, N\geq  1,
\ena
and
\bea \label{epsilon sequence}
0 \to W^{M-2,N+1}_{m-1,n} @>\epsilon^{} >> W^{M,N}_{m,n} @>U>>
W^{M-1,N}_{m,n} \to 0,\qquad M\geq  1, N\geq  -1,
\ena
are exact.
\end{corollary}

By Corollary \ref{short exact sequences} we have the following
recursion relations for characters:
\begin{eqnarray}
{\rm ch} W^{M,N}(q,z_1,z_2)=qz_2\;{\rm ch}
W^{M+1,N-2}(q,q^{-1}z_1,q^2z_2)+{\rm ch} W^{M,N-1}(q,z_1,qz_2),\nonumber\\
{\rm ch} W^{M,N}(q,z_1,z_2)=qz_1\; {\rm ch}
 W^{M-2,N+1}(q,q^2z_1,q^{-1}z_1)+ {\rm ch} W^{M-1,N}(q,qz_1,z_2).\label{rec}
\end{eqnarray}
It can be verified, using the $q$-Pascal identities
\bean
\qbin[m;n]=q^n\qbin[m-1;n]+\qbin[m-1;n-1], \qquad 
\qbin[m;n]=\qbin[m-1;n]+q^{m-n}\qbin[m-1;n-1],
\eean
that the character \Ref{coinchar} satisfies the recursion relations \Ref{rec}.

\subsection{The $fh$-basis}\label{fh basis}

Recall that the spaces $L_{1,0}^{(N)}({\frak n})$ have $ef$ bases given by
\Ref{spanning}. However, we take a quotient with respect to action of $e_i$,
$h_j,f_j$, $i\geq  0$ $j\geq  N$. So, it is more natural to have a basis
in terms of the remaining generators, $h_1,\dots h_{N-1},
f_1,\dots f_{N-1}$. In this section we describe such a basis, which we call
the $fh$-basis.

We are interested in $W^{0,N}$, and $W^{1,N}$ (see Section \ref{l=1}).
We construct the $fh$-bases in these spaces inductively, by using
the short exact sequences \Ref{phi sequence}, \Ref{epsilon sequence}.
In fact, we have to consider spaces $W^{-1,N}$, too, in order to close
the induction steps.

Let ${\cal C}^{0,N}$ be the set of all
$({\bf a},{\bf b})=(a_{N-1},\dots,a_1,a_0; b_{N-1},\dots,b_1)
\in\{0,1\}^{2N-1}$ such that
\bea
\label{triangle relations}
\qquad\quad b_{i+1}+a_{i+1}+a_i\leq  1,\qquad b_{i+1}+b_i+a_i\leq  1,
\ena
\bea
\label{right neighbor}
&&\hbox{for each $i$ such that $b_i\not=0$ there exists $j>i$}\notag \\
&&\hbox{such that $a_j\not=0$ and $b_s=0$ for $i+1\leq  s\leq  j$},
\ena
and
\bean
a_0=0.
\eean

For $({\bf a},{\bf b})\in {\cal C}^{0,N}$, define
$\tau^{(0,N)}({\bf a},{\bf b})\in W$
to be
\bean
\tau^{(0,N)}({\bf a},{\bf b})=\bar f_{N-1}^{a_{N-1}}\bar h_{N-1}^{b_{N-1}}
\dots \bar f_2^{a_2}\bar h_2^{b_2}\bar f_1^{a_1}\bar h_1^{b_1}\bar v.
\eean
Note that $\{\bar f_i, \bar h_j\}$ form a commutative algebra, so we
can write them in any order.

It is convenient to write elements $({\bf a},{\bf b})\in{\cal C}^{0,N}$ as follows.
\bean
\left(\begin{array}{llllllll}
  & b_{N-1} && \cdots& b_2 && b_1 &\\
a_{N-1}&& & \cdots &&a_1 && a_0 =0\end{array}\right).
\eean
Then the map $\tau$ can be read off and relations
\Ref{triangle relations} become ``triangle relations''.

Similarly, let ${\cal C}^{1,N}$ be the set of all $({\bf a},{\bf b})$
satisfying the conditions \Ref{triangle relations}, \Ref{right
neighbor} and that
\bea
\label{leftmost}
&&\hbox{there exists $j$ such that $a_j\not=0$ and $b_s=0$ for $s\leq  j$}.
\ena

For any $({\bf a},{\bf b})\in {\cal C}^{1,N}$, define $\tau^{(1,N)}
({\bf a},{\bf b})\in W$
to be
\bean
\tau^{(1,N)}({\bf a},{\bf b})=
\bar h_N^{b_N}\bar f_{N-1}^{a_{N}}\bar h_{N-1}^{b_{N-1}}\dots\bar f_2^{a_3}
\bar h_2^{b_2}\bar f_1^{a_2}\bar h_1^{b_1}\bar f_0^{a_1}\bar e_1^{1-a_0}\bar v.
\eean

We write $({\bf a},{\bf b})\in{\cal C}^{1,N}$ as 
\bean
\left(\begin{array}{llllllllll}
b_N &&  b_{N-1} &\cdots& b_2 && b_1 &&&\\
&a_{N}&& \cdots &&a_2&&a_1 &&a_0\end{array}\right),
\eean

Finally, set ${\cal C}^{-1,N}={\cal C}^{1,N-2}$. 

For $({\bf a},{\bf b})\in {\cal C}^{-1,N}$, define
$\tau^{(-1,N)}({\bf a},{\bf b})\in W$ to be
\bean
\tau^{(-1,N)}({\bf a},{\bf b})=
\bar f_{N-1}^{a_{N-2}}\bar f_{N-2}^{a_{N-3}}\bar h_{N-2}^{b_{N-2}}\dots \bar f_2^{a_1}\bar h_2^{b_2}\bar f_1^{a_0}\bar h_1^{b_1}\bar v.
\eean

We write $({\bf a},{\bf b})\in{\cal C}^{-1,N}$ as 
\bean
\left(\begin{array}{llllllllll}
&&b_{N-2}&\cdots &&b_3& & b_2 && b_1\\
a_{N-2} &a_{N-3}&& \cdots&a_2& &a_1 &&a_0 &\end{array}\right),
\eean

\begin{example} We have
\bean
W^{-1,1}=W^{1,-1}=W^{-1,0}=W^{0,-1}=0, \qquad W^{0,0}=W^{0,1}=W^{1,0}=\C\bar v,
\eean
\bean
W^{-1,2}=\C\bar f_1\bar v,\qquad W^{1,1}=\langle \bar v,\bar f_0\bar e_1\bar v\rangle,\qquad W^{0,2}=\langle \bar v,\bar f_1 \bar v\rangle.
\eean
\end{example}

\begin{theorem}\label{induction theorem} 
For $M=-1,0,1$, $N=-1,0,1,2,\dots$, $M+N\geq  1$, the vectors 
\bean
\{\tau^{(M,N)}({\bf a},{\bf b})\;|\;({\bf a},{\bf b})\in {\cal C}^{M,N}\}
\eean
form a basis in $W^{M,N}$, and the dimension of $W^{M,N}$ is $2^{M+N-1}$.
\end{theorem}
\begin{proof}
The proof is by induction on $N$. First, by inspection we check that
$W^{-1,1}=W^{1,-1}=0$, and  $W^{0,0}=v$, 
is consistent with the statement of the Theorem.

Next, assume that 
\bean
&\{\tau^{(-1,N_0)}({\bf a},{\bf b})\;|\;({\bf a},{\bf b})\in 
{\cal C}^{-1,N_0}\}{\rm \;\;is\;\; a\;\; basis\;\; of\;\;} W^{-1,N_0},&\\
&\{\tau^{(0,N_0-1)}({\bf a},{\bf b})\;|\;({\bf a},{\bf b})\in 
{\cal C}^{0,N_0-1}\}{\rm \;\;is\;\; a\;\; basis\;\; of\;\;}W^{0,N_0-1},&\\  
&\{\tau^{(1,N_0-2)}({\bf a},{\bf b})\;|\;({\bf a},{\bf b})\in 
{\cal C}^{1,N_0-2}\}{\rm \;\;is\;\; a\;\; basis\;\; of\;\;}W^{1,N_0-2}.& 
\eean
Let us prove that $\{\tau^{(-1,N_0+1)}({\bf a},{\bf b})\;|\;({\bf
a},{\bf b})\in {\cal C}^{-1,N_0+1}\}$ is a basis of $W^{-1,N_0+1}$. We
use the sequences \Ref{phi sequence} with $M=-1$, $N=N_0+1$.

We show the following:
\par\noindent
(i) there exists an injection
$\tilde\varphi:{\cal C}^{0,N_0-1}\to {\cal C}^{-1,N_0+1}$
satisfying
\bean
\varphi(\tau^{(0,N_0-1)}({\bf a},{\bf b}))=\tau^{(-1,N_0+1)}(\tilde\varphi({\bf a},{\bf b}));
\eean
\par\noindent
(ii) there exists a bijection
$\tilde T:{\cal C}^{-1,N_0+1}\slash{\rm Im}\,\tilde\varphi
\rightarrow{\cal C}^{-1,N_0}$ satisfying
\bean
T(\tau^{(-1,N_0+1)}({\bf a},{\bf b}))=\tau^{(-1,N_0)}(\tilde T({\bf a},
{\bf b})).
\eean
In fact, the map (i) is given by
\bean
\tilde\varphi(a_{N_0-2}\dots,a_0;b_{N_0-2}\ldots,b_1)
=(a_{N_0-2}\dots,a_0,1;b_{N_0-2}\ldots,b_1,0).
\eean
Then we have,
\bean
{\cal C}^{-1,N_0+1}\slash{\rm Im}=\{({\bf a},{\bf b})\in{\cal C}^{-1,N_0+1};
a_0=b_1=0\}.
\eean
The map (ii) is given by
\bean
\tilde T(a_{N_0-1},\ldots,a_1,0;b_{N_0-1},\ldots,b_2,0)
=(a_{N_0-1},\ldots,a_1;b_{N_0-1},\ldots,b_2).
\eean

Similarly, we prove that $\{\tau^{(0,N_0)}({\bf a},{\bf b})\;|\;({\bf
a},{\bf b})\in {\cal C}^{0,N_0}\}$ is a basis of $W^{0,N_0}$ using the
sequences \Ref{phi sequence} with $M=0, N=N_0$. Finally, we prove that
$\{\tau^{(1,N_0-1)}({\bf a},{\bf b})\;|\;({\bf a},{\bf b})\in {\cal
C}^{1,N_0-1}\}$ is a basis of $W^{1,N_0-1}$ using the sequences
\Ref{epsilon sequence} with $M=1, N=N_0-1$.\qed
\end{proof}

\begin{corollary}\label{l=0 basis}
The vectors 
\bean
\{f_{N-1}^{a_{N-1}}h_{N-1}^{b_{N-1}}\dots f_2^{a_2}h_2^{b_2}f_1^{a_1}
h_1^{b_1}v\;|\;({\bf a},{\bf b})\in{\cal C}^{0,N}\}
\eean
form a basis in $L_{1,0}^{(N)}({\frak n})$.
\end{corollary}
\begin{proof}
It follows from the remark after Lemma~\ref{deformation}.\qed
\end{proof}

\subsection{The case $l=1$}\label{l=1}
In this section we describe $ef$- and $fh$- bases for
$L_{1,1}^{(N)}({\frak n})$, and compute the corresponding character.

As shown in \cite{FM}, there exists an isomorphism of linear spaces
\bean
D^{1/2}: L_{1,0} \to L_{1,1},
\eean
defined by
\bean
\lefteqn{D^{1/2}(\sigma(f_{i_1}\dots f_{i_a}h_{j_1}\dots h_{j_b}
e_{s_1}\dots e_{s_c}v_{1,0}))=}\\
&&
\sigma(f_{i_1+1}\dots f_{i_a+1}h_{j_1}\dots h_{j_b}e_{s_1-1}\dots 
e_{s_c-1})f_0v_{1,1}.
\eean

Let us define the space of coinvariants
\bean
L_{k,l}^{M,N}({\frak n})=L_{k,l}/\mathfrak{sl}_2^{(M,N)}({\frak n})L_{k,l},
\eean
where we used the notation $\mathfrak{sl}_2^{(M,N)}({\frak n})$ for the subalgebra
of $\mathfrak{sl}_2\otimes\C[t]$ generated by $e_i$ $(i\geq  M)$ and $f_i$ $(i\geq  N)$.
Note that $L^{(N)}_{k,l}({\frak n})=L^{(0,N)}_{k,l}({\frak n})$.

The map $D^{1/2}$ maps $L^{1,N-1}_{1,0}({\frak n})$
to $L_{1,1}^{(N)}({\frak n})$.

\begin{theorem}\label{f-h l=1 basis}
The vectors 
\bean
\{f_{N-1}^{a_{N-1}} h_{N-1}^{b_{N-1}}\dots f_2^{a_2}h_2^{b_2}
f_1^{a_1}h_1^{b_1}f_0^{a_0}v\;|\;({\bf a},{\bf b})\in{\cal C}^{1,N}\}
\eean
form a basis in $L_{1,1}^{(N)}({\frak n})$.
\end{theorem}
\begin{proof}
The $fh$-basis of $W^{1,N-1}$ is described by Theorem \ref{induction
theorem}.  By the argument similar to the proof of Corollary \ref{l=0
basis}, the same vectors form a basis in $L_{1,0}^{1,N-1}({\frak n})$. Theorem
\ref{f-h l=1 basis} is proved by application of the map $D^{1/2}$ to
this basis.\qed
\end{proof}

Similarly, applying the map $D^{1/2}$ to the basis of Theorem
\ref{coinbasis}, we obtain
\begin{theorem}\label{ef l=1 basis}
The vectors
\bean
\{e_{2m-2-n+\alpha_1}\dots e_{2-n+\alpha_{m-1}}
e_{-n+\alpha_m}f_{2n+\beta_1}\dots f_{4+\beta_{n-1}} f_{2+\beta_n}f_0v_1\},
\eean
\bean
n+2-2m>\alpha_1\geq \alpha_2\geq \dots \geq \alpha_m\geq  0,\\
N+m-2n>\beta_1\geq  \beta_2\geq  \dots \geq  \beta_n\geq  0,
\eean
form a basis in $L_{1,1}^{(N)}({\frak n})$.
\end{theorem}

Now, we compute the character of $L_{1,1}^{(N)}({\frak n})$. Using the map
$D^{1/2}$ once again, we obtain
\begin{lemma}
\bea
\label{D-relation on characters}
{\rm ch}_{q,z}L_{1,1}^{(N)}({\frak n})
={\rm ch}_{q,z_1,z_2}W^{1,N-1}|_{z_1=q^{-1}z^2,z_2=qz^{-2}}.
\ena
\end{lemma}

\begin{theorem}\label{l=1 character}
\bean
{\rm ch}_{q,z}L_{1,1}^{(N)}({\frak n})
=\sum_{s\geq 0}q^{s(s-1)}\qbin[N;2s-1]z^{-2s-1}.
\eean
\end{theorem}

\begin{remark}
Taking the limit $N\to\infty$ of the characters
${\rm ch}_{q,z}L_{1,l}^{(N)}({\frak n})$
$(l=0,1)$, given by Theorems \ref{charform} and \ref{l=1 character}, we obtain
\bean
{\rm lim}_{N\rightarrow\infty}
{\rm ch}_{q,z}L_{1,l}^{(N)}({\frak n})
=\sum_{s}q^{s(s-l)}\frac{1}{(q)_{2s-l}}\,z^{-2s-l},
\eean
where $(q)_m=(1-q)(1-q^2)\dots(1-q^m)$.
At $z=1$ the above characters
coincide with the characters of Virasoro modules in the minimal conformal
field theory ${\cal M}(3,5)$ with central charge $-3/5$.
Namely, the character for $l=0$ is the representation
$\phi_{1,2}$ with conformal dimension $-1/20$ and the character for
$l=1$ is the representation $\phi_{1,3}$ with conformal dimension $1/5$,
cf. \cite{KKMM}.
\end{remark}

\section{Appendix}
\subsection{The Verlinde rule}
Here we prove Theorem \ref{Ver} using the standard Verlinde rule
proved in (\cite{TUY}). 
For simplicity we work in the
$\mathfrak{sl}_2$ case, but this method works for any simple Lie algebra.
 
Let us first recall the standard statement of the Verlinde rule (Theorem
\ref{cit} below).

To simplify the notation for the coinvariant spaces we will write $M/A$
instead of $M/AM$ for an algebra $A$ and an $A$--module $M$ if it does not
lead to confusion.

Let $z_1, \dots, z_N$ be distinct points on $\proj$, $U$ the
one--dimensional complex variety $\proj\setminus \{z_1, \dots, z_N\}$
and $A$ the ring of regular functions on $U$, i.e., meromorphic
functions on $\proj$ with possible poles at $z_1,\ldots,z_N$.

Let $t$ be a coordinate on $\proj \setminus \infty \cong \CC$.
We choose a local coordinate $t_i$ at $z_i$ as follows:
\bea
t_i=
\cases{
t-z_i&if\quad$z_i\not=\infty$;\cr
t^{-1}&if\quad$z_i=\infty$.}
\nonumber
\ena

We have the inclusion of $\slg$--valued functions
$$\slg\otimes A \hookrightarrow \bigoplus_{i=1}^N
\slg\otimes\C[t_i^{-1},t_i]],$$
given by the Laurent expansions at the points $z_i$ in $t_i$. 

\def \haz {\hat{\mathfrak A}(z_1, \dots, z_N)}

The Lie algebra in the right hand side has a canonical cocycle which is
 a sum of canonical cocycles of of the summands. We denote the
 corresponding central extension by $\haz$. We have
$$
[\sum_{i=1}^NX_i\otimes f(t_i),\sum_{i=1}^NY_i\otimes g(t_i)]
=\sum_{i=1}^N[X_i,Y_i]\otimes f(t_i)g(t_i)
+c\,(X,Y)\,\sum_{i=1}^N{{\ }\atop{\mbox{\Large
Res}\atop{t_i=0}}}\left(g(t_i)df(t_i)\right),
$$
where $X_i,Y_i\in \mathfrak{sl}_2$, $f(t),g(t)$ are Laurent polynomials, $c$ is the central element and $(X,Y)$ is the Killing form.

Note that the restriction of this cocycle to $\slg\otimes A$ is
trivial, therefore we have the inclusion of Lie algebras
$$\slg \otimes A \hookrightarrow \haz.$$

\def\slgi{\widehat{{\mathfrak{sl}}}_2^{(i)}}

Now fix the level $k\in\Z_{>0}$ and consider a sequence
$\{\rho_1,\dots,\rho_N\}$ of integrable irreducible $\widehat{\mathfrak{sl}}_2$--modules
at level $k$. Namely, $\rho_i=L_{k,l_i}$ for some $0\leq l_i\leq k$ in our
notation. We can treat each
$\rho_i$ as a representation of
\bea\notag
\slgi\buildrel{\hbox{\rm def}}\over=
\slg\otimes\CC[t_i^{-1},t_i]]\oplus\CC c,
\ena
and the tensor product $\rho_1\otimes\dots\otimes\rho_N$ as a representation of
$\haz$. 

Denote by $\left<\rho_1,\ldots,\rho_N\right>$
the coefficient of the unit $\pi_0$ in the product
$\pi_{l_1}\cdots\pi_{l_N}$  in the Verlinde algebra $\Vk$
(see (\ref{VERLINDE})).

\begin{theorem} \label{cit}(Corollary 6.2.5 in \cite{TUY}.)
The dimension of the coinvariant space 
$(\rho_1\otimes\dots\otimes\rho_N )/ (\slg\otimes A)$ is equal to 
$\left<\rho_1,\ldots,\rho_N\right>$.
\end{theorem}

\def\slginf{\widehat{{\mathfrak{sl}}}_2^{(\infty)}}

\def\hazm{\widehat{\mathfrak{A}}(z_1, \dots, z_{N-1})}

We suppose that $z_N = \infty$. We denote by $\rho$ the representation
$\rho_N$. Then $\rho$ is a highest
weight module of 
\bea 
\slginf={\mathfrak{sl}}_2\otimes\C[t,\frac1t]]\oplus\C c. \notag
\ena 

In what
follows we use two different central extensions of the algebra
$\slg\otimes A$: One is obtained from the embedding in $\slginf$, and
the other in $\hazm$. The corresponding cocycles differ by a minus
sign.

Let $B_L$ $(L\ge0)$ be the Lie subalgebra $\slg\otimes \C[t]\cdot
\prod\limits_{i=1}^{N-1} t_i^L$ of $\slginf$. Note that
$B_1=\mathfrak{sl}_2({\cal Z};0)$,
where ${\cal Z}=(z_1,\dots,z_{N-1})$.

\begin{lemma}\label{finite dimensions}
The space of coinvariants
$$I_L = \rho/B_L$$
is finite--dimensional.
\end{lemma}

\begin{proof}
This is a special case of Theorem~3.1.3 with $(N-1)L$ non--distinct
points.\qed
\end{proof}

\def \J {{\mathcal J}}

We have the natural projection $I_L\to I_{L-1}$. Let $\J$ be the limit of
the injective system of dual spaces $\cdots \to I_L^* \to I_{L+1}^* \to
\cdots$. Here we consider the dual action of 
$a\in {\mathfrak{sl}}_2\otimes\C[t,\frac1t]]$ on $\rho^*={\rm Hom}(\rho,\CC)$ by
\bea\notag
\left<a\cdot v^*,v\right>+\left<v^*,a\cdot v\right>=0.
\ena

\begin{lemma}
There is an action of the Lie algebra $\hazm$ on $\J$ at level $k$.
\end{lemma}

\begin{proof}

For an element $a \in \slg\otimes A$, 
there exists an integer $L$ such that $[a, B_j] \subset B_{j-L}$ for any $j>L$.
Therefore, the element $a$ defines a map $I_j \to I_{j-L}$ and a dual map
$I_{j-L}^* \to I_L^*$. Note that the action of $\slginf$
on $\rho^*$ is of level $-k$. Therefore, we have an action of the central
extension of $\slg\otimes A\subset\slginf$ at level $-k$ on the limit $\J$.

Introduce the topology in the central extension of $\slg\otimes
A\subset\hazm$ where open sets are $B_L$. The completion in this topology
is exactly $\hazm$.  Note that for any vector $v\in \J$ some $B_L$
acts trivially on $v$. Therefore, we have the action of this
completion on $\J$. Since as we noted the cocycle of $\slg\otimes A$ in $\haz$
is trivial, the cocycle of $\slg\otimes A$ in $\hazm$ is equal to the
negative of the cocycle in $\slginf$. Therefore, the level of the
representation of $\hazm$ obtained by the completion is $k$.\qed
\end{proof}

A representation of the Lie algebra $\hazm$ is called {\em integrable} if it
is a direct sum of finite dimensional modules with respect to
each of the ${\mathfrak{sl}}_2$ subalgebras generated by 
$e\otimes t_i^{-j},f\otimes t_i^j$ $(i=1,\ldots,N-1;j=0,1)$.

\begin{lemma}
The $\hazm$ module $\J$ is integrable.
\end{lemma}

\begin{proof}
Note that the subalgebra $\slg\otimes\CC[t_i]$ belongs to $B_0$. The
action on $\J$ of the ${\mathfrak{sl}}_2$ subalgebra of
$\slg\otimes\CC[t_i]$ generated by $e\otimes1$ and $f\otimes1$ is integrable
because $\J$ is the inductive limit of the finite dimensional spaces
$(\rho/B_L)^*$ on which $B_0$ acts.

In order to see the integrability with respect to the ${\mathfrak{sl}}_2$
subalgebra of
\bea
(\CC e\otimes t_i^{-1}\CC[t_i])\oplus(\CC h\otimes\CC[t_i])
\oplus(\CC f\otimes t_i\CC[t_i])\label{TWIST}
\ena
that is generated by $e\otimes t_i^{-1}$ and $f\otimes t_i$,
we define the subalgebras
\bea
C_L=
(\CC e\otimes\CC[t]\cdot\prod_{i=1}^{N-1} t_i^{L-1})
\oplus(\CC h\otimes\CC[t]\cdot\prod_{i=1}^{N-1} t_i^L)
\oplus(\CC f\otimes\CC[t]\cdot\prod_{i=1}^{N-1} t_i^{L+1}).\nonumber
\ena
The subalgebra (\ref{TWIST}) belongs to the completion of
$C_0$. Since $C_L\subset B_{L-1}\subset C_{L-2}$, the module $\J$ is
also the inductive limit of the finite dimensional spaces $(\rho/C_L)^*$
on which the completion of $C_0$ acts. Therefore, the integrability follows.
\qed
\end{proof}

We say that a representation of $\hazm$ belongs to the category ${\mathcal O}$
if the action of the subalgebra generated by the elements
$e\otimes1,f\otimes t_i\in\slgi$
$(i=1,\ldots,N-1)$ is locally nilpotent, i.e., nilpotent on
each vector of the representation.

Our next goal is to prove that the representation $\J$ belongs to the
category ${\mathcal O}$. We need a few standard lemmas from
the theory of Lie algebra.

\begin{lemma}\label{SOLV}
Let $\mathfrak{n}$ be a solvable Lie algebra. Then, any irreducible
representation of $\mathfrak{n}$ is one dimensional.
\end{lemma}

\begin{lemma}\label{NILP}
Let $\mathfrak{n}$ be an ideal of a Lie algebra $\mathfrak{a}$.
Let $V$ be a representation of $\mathfrak{a}$,
$\psi$ an element of the dual space ${\mathfrak{n}}^*$, and
\bea\notag
V_\psi=\{v\in V; Xv=\psi(X)v\hbox{ for all $X\in{\mathfrak{n}}$}\}.\nonumber
\ena
Then, $V_\psi$ is $\mathfrak{a}$--invariant.
\end{lemma}

Using these lemmas we can prove the following
\begin{lemma}\label{BORYA}
Let $\mathfrak{n}$ be a nilpotent subalgebra of a Lie algebra $\mathfrak{a}$
such that
\bea\notag
[{\mathfrak{a}},{\mathfrak{n}}]={\mathfrak{n}}.
\ena
Let $V$ be a finite dimensional representation of $\mathfrak{a}$.
Then, the action of $\mathfrak{n}$ on $V$ is nilpotent.
Moreover, $V$ has a filtration by $\mathfrak{a}$--invariant subspaces
such that the action of $\mathfrak{n}$ on the gradation is trivial.
\end{lemma}
\begin{proof}
We prove the statement by induction on ${\rm dim}\,V$.
If ${\rm dim}\,V=0$, the statement is obvious.
If ${\rm dim}\,V\ge1$, because of Lemma \ref{SOLV},
we have a subspace $\CC v$ $(v\not=0)$ invariant with respect to
$\mathfrak{n}$. Define $\psi\in{\mathfrak{n}}^*$ by $Xv=\psi(X)v$.
We will show that $\psi=0$. By Lemma \ref{NILP}, $V_\psi$
is $\mathfrak{a}$--invariant, and therefore, the action of
$[{\mathfrak{a}},{\mathfrak{n}}]$ on $V_\psi$ is trivial. Since
$[{\mathfrak{a}},{\mathfrak{n}}]={\mathfrak{n}}$, we have $\psi=0$.
In other words, the subspace
$V_0=\{v\in V;Xv=0\hbox{ for all $X\in{\mathfrak{n}}$}\}$
is nonzero and $\mathfrak{a}$--invariant.
By the induction hypothesis, the action of $\mathfrak{n}$ on the $\mathfrak{a}$
module $V/V_0$ is nilpotent, and therefore the action on $V$ itself, too.
In the proof, we have also constructed a filtration which satisfies
the requirement of the lemma.\qed
\end{proof}

{}From these preparations follows
\begin{lemma}
The action on $(\rho/B_L)^*$ of
the subalgebra generated by the $2(N-1)$ elements
$e\otimes1,f\otimes t_i\in\slgi$
$(i=1,\ldots,N-1)$ is nilpotent.
\end{lemma}
\begin{proof}
Since $[\slgi,\widehat{\mathfrak{sl}}_2^{(j)}]=0$
if $i\not=j$,
it is enough to prove the lemma for a single $i$. We use Lemma \ref{BORYA}
by setting ${\mathfrak{a}}=\mathfrak{sl}_2\otimes\CC[t_i]$,
${\mathfrak{n}}=\mathfrak{sl}_2\otimes t_i\CC[t_i]$ and $V=(\rho/B_L)^*$.
The statement follows immediately.\qed
\end{proof}

Therefore, we have
\begin{lemma}
The representation $\J$ of $\hazm$ belongs to the category ${\mathcal O}$.
\end{lemma}

We can apply the following standard lemma to $\J$.
\begin{lemma}
If an integrable representation of $\hazm$ at level $k$ belongs to the
category ${\mathcal O}$, then it is a direct sum of
tensor product $L_1\otimes\cdots\otimes L_{N-1}$ where $L_i$ are integrable
irreducible representations of $\slgi$ at level $k$.
\end{lemma}

Note that there is an isomorphisms of Lie algebras
\bea\notag
\iota&:&\slg\otimes\CC[t]/B_1\rightarrow
\underbrace{\slg\oplus\cdots\oplus\slg}_{N-1}
\ena
induced from the mappings
\bea
\slg\otimes\CC[t]\rightarrow
\slg\otimes\CC[t_1]\oplus\cdots\oplus
\slg\otimes\CC[t_{N-1}]\rightarrow
{\mathfrak{sl}}_2\oplus\cdots\oplus{\mathfrak{sl}}_2,
\label{EVAL}
\ena
where the last arrow is given by the evaluation homomorphism at
$t_i=0$ $(i=1,\ldots,N-1)$ that is
$\iota(X\otimes f(t))=(X\otimes f(z_1))\oplus\cdots\oplus
(X\otimes f(z_{N-1}))$.
Since the action of $B_1$ on $\rho/B_1$ is trivial, there is an action of
$\slg\otimes\CC[t]/B_1\simeq\slg\oplus\cdots\oplus\slg$
on $(\rho/B_1)^*$. 

\begin{lemma} \label{res}
If an irreducible representation $\nu_1\otimes\dots\otimes \nu_{N-1}$
of $\slg\oplus \dots \oplus \slg$ appears in the decomposition of $\slg\oplus
\dots \oplus \slg$ module $(\rho/B_1)^*$ then $\dim \nu_j \leq k+1$.

\end{lemma}

\begin{proof}
Note that the completion of $B_1$ in $\hazm$ is equal to
$$
\widehat B_1=
\slg\otimes t_1\CC[t_1]\oplus \dots \oplus \slg\otimes t_{N-1}\CC[t_{N-1}].
$$
The space $(\rho/B_1)^*$ is the space of $\widehat B_1$--invariants 
in the $\hazm$--module $\J$. We know that $\J$ is a direct sum of
products $L_1\otimes\dots \otimes L_{N-1}$ of integrable modules and that
the space of $\slg \otimes t_i\CC[t_i]$ --invariants in integrable module
at level $k$ is an irreducible $\slg$--module of dimension at most $k+1$.
Therefore, we have the statement of the lemma.\qed
\end{proof}

Let $\nu_i$ $(i=1,\ldots,N-1)$ be irreducible finite--dimensional
representations of $\slg$. We denote by $M(\nu_i)$ the
$\slgi$--module at level $k$ induced from the
evaluation representation $\nu_i$ of $\slg\otimes \CC[t_i]$ at $t_i=0$:
$$
M(\nu_i)={\rm Ind}^{\slgi}_{{\mathfrak{sl}}_2\otimes \CC[t_i]}\nu_i
\buildrel{\rm def}\over=
U(\slgi)/U(\slgi)(c-k)\bigotimes_{{\mathfrak{sl}}_2\otimes \CC[t_i]}\nu_i.
$$
Then, the tensor product $M(\nu_1)\otimes\dots\otimes M(\nu_{N-1})\otimes
\rho$ 
can be considered as a representation of $\hat{\mathfrak A}(z_1, \dots,
z_{N-1}, \infty)$ and therefore of $\slg\otimes A$.

\begin{lemma}\label{iso1}
There is a non--degenerate coupling between
$$
\left(M(\nu_1)\otimes\dots\otimes
M(\nu_{N-1})\otimes\rho\right)/\slg\otimes A
$$
and
$$
{\rm Hom}_{\slg\oplus\dots\oplus\slg}
\left(\nu_1\otimes\dots\otimes\nu_{N-1},(\rho/B_1)^*\right).
$$
\end{lemma}

\begin{proof}
Clearly, 
$$\left(M(\nu_1)\otimes\dots\otimes M(\nu_{N-1})\otimes \rho\right)/
\slg\otimes A \cong M(\nu_1)\otimes\dots\otimes
M(\nu_{N-1})\bigotimes_{\slg\otimes A} \rho.$$

There is an action of $\slg\otimes\CC[t]$
on the tensor product $\nu_1\otimes\cdots\otimes\nu_{N-1}$
through (\ref{EVAL}). Clearly, we have
$$M(\nu_1)\otimes\dots\otimes M(\nu_{N-1}) \cong {\rm Ind}^{\slg\otimes
A}_{\slg\otimes \CC[t]} \nu_1\otimes\dots\otimes\nu_{N-1}.$$
Therefore, we have
$$M(\nu_1)\otimes\dots\otimes M(\nu_{N-1}) \bigotimes_{\slg\otimes A }
\rho
\cong 
\nu_1\otimes\dots\otimes \nu_{N-1} \bigotimes_{\slg\otimes \CC[t] } \rho.$$ 

As $B_1\subset \slg\otimes\CC[t]$ acts on $\nu_1\otimes\dots\otimes
\nu_{N-1}$
by zero, we get 
$$\nu_1\otimes\dots\otimes \nu_{N-1} \bigotimes_{\slg\otimes \CC[t] } \rho\cong
\nu_1\otimes\dots\otimes \nu_{N-1} \bigotimes_{\slg\otimes \CC[t]/B_1 }
\rho/B_1,$$
which implies the lemma.\qed
\end{proof}

Recall that $\rho_i=L_{k,l_i}$.

\begin{theorem} \label{iso2}
We have an equality of dimensions,
$$
{\rm dim}\,\left(\rho_1\otimes\dots\otimes
\rho_{N-1}\otimes \rho\right)/ \slg\otimes A
={\rm dim}\,{\rm Hom}_{\slg\oplus\dots\oplus\slg}
\left( \pi_{l_1}\otimes\dots\otimes\pi_{l_{N-1}}, (\rho/B_1)^*\right).
$$
\end{theorem} 

\begin{proof}
 Let us set $\nu_i = \pi_{l_i}$. 
 By Lemma \ref{iso1} it is enough to prove that the natural surjection
\bea\label{ISOMOR}
\left(M(\nu_1)\otimes\dots \otimes M(\nu_{N-1})\otimes \rho\right)/
\mathfrak{sl}_2\otimes A
\to
\left(\rho_1\otimes\dots\otimes \rho_{N-1}\otimes \rho \right)/
\mathfrak{sl}_2\otimes A           
\ena
is an isomorphism.

Consider the exact sequence
$$M(\nu_1') \to M(\nu_1) \to \rho_1 \to 0.$$

A standard computation with the Weyl group shows that $\dim \nu_1' =
2k+3-l_1$.

We have the exact sequence of coinvariants
\begin{eqnarray*}
\left(M(\nu_1')\otimes\dots \otimes M(\nu_{N-1})\otimes \rho \right)/
\slg\otimes A
\to \\
\to \left(M(\nu_1)\otimes\dots \otimes M(\nu_{N-1})\otimes \rho \right)/
\slg\otimes
A \to \\
\to \left(\rho_1\otimes\dots \otimes M(\nu_{N-1})\otimes \rho \right)/
\slg\otimes A
\to 0.
\end{eqnarray*}

Since $\dim \nu_1' > k+1$, by Lemma \ref{res}, the first term is zero.
Therefore, the second map is an
isomorphism. Repeating this procedure for $\nu_2, \dots, \nu_{N-1}$ we
obtain that \Ref{ISOMOR} is an isomorphism. (Cf. \cite{Fi} for an
alternative proof.) \qed
\end{proof}

\begin{corollary} \label{c1}
Let $z_1, \dots, z_{N-1}$ be distinct complex numbers. Then

(i) (case of the bigger coinvariants)
$$ \dim L_{k,l}/\mathfrak{sl}_2\otimes \CC[t]\prod_{i=1}^{N-1} (t-z_i) = 
\sum_{0\leq l_1, \dots,
l_{N-1}\leq k} \left< \pi_{l_1}, \dots , \pi_{l_{N-1}}, \pi_l\right> \dim
(\pi_{l_1}\otimes\dots\otimes \pi_{l_{N-1}}),$$

(ii) (case of the smaller coinvariants)
$$ \dim L_{k,l}/\left(
(\CC f \oplus \CC h)\otimes \CC[t]\prod_{i=1}^{N-1} (t-z_i) \oplus
\CC e \otimes \CC[t]
\right) = \sum_{0\leq l_1, \dots,
l_{N-1}\leq k}
\left< \pi_{l_1}, \dots, \pi_{l_{N-1}}, \pi_l\right>.$$
\end{corollary}

\begin{proof}
The statement (i) follows from Theorems \ref{iso2} and \ref{cit}.
The statement (ii) follows from (i) because the quotient of
an irreducible $\slg\oplus\dots\oplus \slg$--module by the 
subalgebra $\CC e\oplus\dots\oplus\CC e$ is one--dimensional space.\qed
\end{proof}

This corollary is clearly equivalent to Theorem \ref{Ver}.


\begin{thebibliography}{KKMM}

\bibitem[F]{F}B. Feigin, Conformal field theory and cohomologies of the Lie
algebra of holomorphic vector fields on a complex curve,
in Proceedings of the ICM at Kyoto, 1990.

\bibitem[FF]{FF}B. Feigin and E. Frenkel, Coinvariants of nilpotent subalgebras
of the Virasoro algebra and partition identities, Adv. Sov. Math.
{\bf 16} (1993), 139-148.

\bibitem[FL]{FL}B. Feigin and S. Loktev, On generalized Kostka polynomials
and quantum Verlinde rule, math.QA/9812093.

\bibitem[FM]{FM}B. Feigin and T. Miwa, Extended vertex operator algebras
and monomial bases, math.QA/9901067.  

\bibitem[FS]{FS}B. Feigin and A. Stoyanovsky, Quasi-particles models for the
representations of Lie algebras and geometry of flag manifold,
hep-th/9308079, RIMS 942; Functional models for the representations of
current algebras and the semi-infinite Schubert cells, {\it
Funct. Anal. Appl.} {\bf 28} (1994), 55--72.

\bibitem[Fi]{Fi}M. Finkelberg, An equivalence of fusion categories,
Geom. Funct. Anal. {\bf 6}(1996)249-267.

\bibitem[KKMM]{KKMM}R. Kedem, T. Klassen, B. McCoy, E. Melzer,
Fermionic sum representations for conformal field theory
characters. {\it Phys. Lett.}  {\bf B 307} (1993), 68--76.

\bibitem[S]{S}G. Segal, Geometric aspect of quantum field theory,
in Proceedings of the ICM at Kyoto, 1990.

\bibitem[St]{St}A. Stoyanovsky, Lie algebra deformation and character formulas,
Func. Anal. Its Appl.,{\bf 32}(1998)66-68.

\bibitem[TUY]{TUY}A. Tsuchiya, K. Ueno and Y. Yamada, Conformal field theory
on the universal family
of stable curves with gauge symmetry, Adv. Stud. Pure Math.,
{\bf 19}(1989)459-466.

\bibitem[V]{V}E. Verlinde, Fusion rules and modular transformations in 2D
conformal field theory, Nuclear Physics {\bf B300}(1988)360-376.

\end{thebibliography}
\end{document}